\documentclass[fleqn,usenatbib]{mnras}

\usepackage{newtxtext,newtxmath}
\usepackage[T1]{fontenc}

\DeclareRobustCommand{\VAN}[3]{#2}
\let\VANthebibliography\thebibliography
\def\thebibliography{\DeclareRobustCommand{\VAN}[3]{##3}\VANthebibliography}

\usepackage{graphicx}
\usepackage{amsmath}
\usepackage{natbib}
\usepackage{longtable}
\usepackage[labelfont=bf,labelsep=period]{caption}  % for longtable
\LTcapwidth=\textwidth  % for longtable

\title[Correlated rotation curves \& mass models]{Consequences of radially correlated rotation curves for galaxy mass models}

\author[H.~Chase et al.]{
Helena Chase$^{1,2}$, 
Diego Dado$^{1,2}$,
Katherine E. Harborne$^{1,2}$ and
Kyle A. Oman$^{1,2}$\thanks{E-mail: kyle.a.oman@durham.ac.uk}
\\
$^{1}$Institute for Computational Cosmology, Physics Department, Durham University, South Road, Durham DH1 3LE, United Kingdom\\
$^{2}$Centre for Extragalactic Astronomy, Physics Department, Durham University, South Road, Durham DH1 3LE, United Kingdom\\
}

\date{Accepted XXX. Received YYY; in original form ZZZ}
\pubyear{\the\year{}}

\begin{document}
\label{firstpage}
\pagerange{\pageref{firstpage}--\pageref{lastpage}}
\maketitle

\begin{abstract}
Consecutive points in rotation curve measurements are correlated with each other, but this is usually ignored when constructing galaxy mass models. We apply a recently proposed data-driven approach to include the characteristic amplitude and scale length of such correlations as `nuisance parameters'. We construct mass models for $134$ galaxies from the SPARC rotation curve compilation with Navarro-Frenk-White (NFW) and pseudo-isothermal sphere (pISO) models for the dark halo. Allowing for correlations in the rotation curves generally improves the goodness of fit for both halo models, often yielding a formally good fit ($\chi^2_\mathrm{r}\approx 1$) and model uncertainties that seem more representative of the constraining power of the data. For both halo models the inference on the typical correlation amplitude and scale length are very similar and physically plausible, $\sim 20\,\mathrm{km}\,\mathrm{s}^{-1}$ and $\sim 5\,\mathrm{kpc}$, respectively. The parametric form that we use to describe the correlations is intentionally simple, and our fitting approach  makes the parameters describing possible correlations prone to `absorbing' other systematic errors, so we regard these estimates as upper limits. Without allowing for correlations we find a statistical preference for the pISO over the NFW model for $88$/$134$ galaxies; this preference essentially disappears when correlations are allowed for. Accounting for correlations in rotation curves when constructing mass models fundamentally affects how they are interpreted, highlighting an important systematic uncertainty that affects evidence for cusps or cores in dark matter haloes.
\end{abstract}


\section{Introduction} \label{sec:intro}

Rotation curves can provide strong constraints on the mass and structure of galaxies. Since early observations these have showed a discrepancy between the visible matter and the inferred dynamical mass \citep[see][for a historical perspective]{2018RvMP...90d5002B}. Mass modelling can reveal the distribution of the dark matter within galaxies. Several dark halo models with differing density profiles have been proposed \citep[e.g.][]{1972ApJ...176....1G,1990ApJ...356..359H,1995ApJ...447L..25B,1996ApJ...462..563N,2014MNRAS.437..415D,2016MNRAS.459.2573R} with the aim of providing the best fit to the observed kinematics within prior constraints.

Several authors \citep[e.g.][]{2001AJ....122.2396D,2006ApJS..165..461K,2008AJ....136.2648D,2019MNRAS.488.5127F,2020ApJS..247...31L} have argued that allowing for centrally constant dark matter density `cores' \citep[][and see \citealp{2010AdAst2010E...5D,2022NatAs...6..897S} for reviews]{1994ApJ...427L...1F,1994Natur.370..629M} can lead to better-fitting mass models than centrally denser dark matter `cusps', especially for low-mass galaxies. However, central density cusps are a generic prediction of N-body simulations of cosmological structure formation in the $\Lambda$CDM cosmogony \citep{1997ApJ...490..493N}.

Mass model fitting is subject to many potential systematic errors arising following uncertain assumptions \citep[e.g.][]{2004ApJ...617.1059R,2007ApJ...657..773V,2016MNRAS.462.3628R,2017MNRAS.466...63P}. This work focuses on one potential source of error: in nearly all prior work, the velocity measurements making up a rotation curve are treated as statistically independent, when in reality they are correlated. Correlation arises in several ways. First and least controversially, the circular velocity is an integral quantity that depends at least on the inner matter distribution (in a spherically symmetric system), or more generally on the entire matter distribution. All points in a rotation curve are therefore at minimum correlated with all points interior to themselves, if the kinematics are gravitationally driven. Instrumental effects, such as beam smearing \citep[e.g.][]{2009A&A...493..871S}, can also introduce local correlations. Physical effects also play a role, for example a spiral arm can induce a locally correlated perturbation to the rotation pattern -- indeed local morphological features often correspond to local features in rotation curves \citep[][but see also \citealp{2025arXiv250803569K}]{2004IAUS..220..233S}.

Given the diversity of origins of correlations in rotation curve data it is prohibitively difficult to explicitly account for all of them in mass models. Instead, \citet{2022RNAAS...6..233P} proposed a data-driven approach. A covariance matrix is included in the mass modelling pipeline and Gaussian process (GP) regression is used to constrain the correlation amplitudes. As a starting point the correlations are assumed to obey a simple analytic form described by a single amplitude and length scale (per galaxy). His work presented NGC~2403 as a case study and showed that the data were consistent with the presence of correlations of $3\,\mathrm{km}\,\mathrm{s}^{-1}$ on scales of $1\,\mathrm{kpc}$. He further showed that failing to account for the possible presence of these correlations can lead to underestimated uncertainties and biased estimates for structural parameters, such as the dark halo mass. \citet{2024MNRAS.532L..48O} applied the same methodology to the case of the Milky~Way and showed that low-amplitude correlations ($\lesssim 10\,\mathrm{km}\,\mathrm{s}^{-1}$) on $\sim 2\,\mathrm{kpc}$ scales are likely to be present and accounting for them leads to a 50~per~cent lower total mass for the Galaxy.

In this work, we fit mass models to the rotation curves of $134$ galaxies from the SPARC\footnote{Spitzer Photometry and Accurate Rotation Curves.} compilation (Sec.~\ref{sec:data}) assuming either a density cusp or core in the dark halo (Sec.~\ref{sec:method}). In each case we also fit models both with and without accounting for possible correlations in the rotation curve data. We investigate the differences between these models with a particular focus on differences in goodness of fit (Sec.~\ref{sec:results}), and discuss changes in the maximum-likelihood values and uncertainty in the model parameters. We comment on how our findings affect the interpretation of rotation curve data and where this differs from the conclusions drawn in prior work (Sec.~\ref{sec:discussion}).

We assume a cosmology with a Hubble-Lema\^{i}tre parameter $H_0=73\,\mathrm{km}\,\mathrm{s}^{-1}\,\mathrm{Mpc}^{-1}$ for consistency with \citet{2016AJ....152..157L}. Virial quantities are defined by a overdensity of $200$ times the critical density for closure $\rho_\mathrm{crit}=3H_0^2/8\pi G$ where $G$ is Newton's constant, i.e. $M_\mathrm{200c}$ is the mass enclosed within a sphere of radius $R_\mathrm{200c}$ within which the mean density is $200\rho_\mathrm{crit}$.

\section{Rotation curve sample} \label{sec:data}

Our sample of rotation curves is drawn from the SPARC \citep{2016AJ....152..157L} compilation of $175$ nearby galaxies with $3.6 \mathrm{{\mu}m}$ surface photometry and H\,\textsc{i} and/or H$\alpha$ rotation curves. It spans a wide range of late-type galaxy properties including morphology, luminosity, and surface brightness. The H\,\textsc{i}/H$\alpha$ measurements aim to trace the circular velocity curves of the galaxies (e.g. corrected for inclination), while the $3.6 \mathrm{{\mu}m}$ band minimizes the uncertainty in the stellar mass-to-light ratio \citep{2014ApJ...788..144M}. In many galaxies the rotation curves extend to large radii and can constrain the structural parameters of the dark halo.

We adopt bulk galaxy properties from the SPARC catalogue including the distance and inclination of each galaxy. To isolate and focus our attention on the leading-order influence of allowing for radial correlations in the rotation curve measurements we hold these parameters fixed at their fiducial values \citep[][table~1]{2016AJ....152..157L} throughout this work, but note that in full detail the mass models parameters will covary with these if more freedom is allowed. We take the $v_\mathrm{rot}$ values and uncertainties tabulated in \citet[][table~2]{2016AJ....152..157L} as the rotation curves of the galaxies and fix the shapes of the stellar disc \& bulge and cold gas contributions to the mass model to $v_\mathrm{bulge}$, $v_\mathrm{disc}$ and $v_\mathrm{gas}$ from the same table\footnote{In that table these are denoted $V_\mathrm{obs}$, $V_\mathrm{bul}$, $V_\mathrm{disk}$ [sic] and $V_\mathrm{gas}$, respectively.}. We fix the normalisation of the gas component to the tabulated value, but allow freedom in the stellar mass-to-light ratios as detailed below (Sec.~\ref{subsec:massmodel}).

\section{Mass model definitions and fitting method} \label{sec:method}

\subsection{Mass model} \label{subsec:massmodel}

We closely follow the methodology of \citet{2022RNAAS...6..233P}. We therefore summarise the fitting procedure emphasising where we have modified it. The rotation curve is decomposed into components as:
\begin{equation}
    v^2_\mathrm{rot}(R) = v^2_\mathrm{gas}(R) + \left(\rule{0em}{3.5mm}\Upsilon_\mathrm{disc} v_\mathrm{disc}(R)\right)^2 + \left(\Upsilon_\mathrm{bulge} v_\mathrm{bulge}(R)\right)^2 + v^2_\mathrm{DM}(R) \; ,
\end{equation}
where $v_\mathrm{gas}(R)$, $v_\mathrm{disc}(R)$, $v_\mathrm{bulge}(R)$ and $v_\mathrm{DM}(R)$ are the velocity contributions as a function of radius, $R$, from the gas, stellar disc, stellar bulge\footnote{\citet{2022RNAAS...6..233P} did not include a bulge component since NGC~2403 does not have one.}, and dark matter halo, respectively, and $\Upsilon_\mathrm{disc}$ and $\Upsilon_\mathrm{bulge}$ are the stellar mass-to-light ratios of the disc and bulge. We adopt the $v_\mathrm{gas}(R)$, $v_\mathrm{disc}(R)$ and $v_\mathrm{bulge}(R)$ profiles tabulated in the SPARC compilation.

We consider two models for the dark halo: the \citet[NFW;][]{1996ApJ...462..563N} profile and the pseudo-isothermal \citep[pISO;][]{1972ApJ...176....1G} profile \citep[only the former was used by][]{2022RNAAS...6..233P}. Although other models are known to be able to provide formally better statistical fits to simulated dark matter haloes and/or galaxy rotation curves \citep[e.g.][]{1995ApJ...447L..25B,2006AJ....132.2685M,2010MNRAS.402...21N,2016MNRAS.457.4340K,2016MNRAS.459.2573R,2017MNRAS.466.1648K,2017A&A...605A..55A,2020MNRAS.497.2393L,2020ApJS..247...31L}, we choose these two because: (i) they are simple; (ii) they have been widely used in previous work, facilitating comparison; and most importantly (iii) they maximise the difference between the two models with very different density profile slopes that bracket the range of plausible halo models. This choice helps to emphasise differences in model fits that could be more subtle had we chosen more similar or flexible models. Our aim is not to find the best mass models but instead to explore the influence of allowing for (or neglecting) correlations in the rotation curve data.

The NFW profile is characterised by a central $\rho \propto R^{-1}$ cusp and an outer slope of $\rho \propto R^{-3}$ and has two degrees of freedom. One possible parametrization is the halo mass, $M_\mathrm{200c}$, and the concentration, $c_\mathrm{NFW}\equiv R_\mathrm{200c}/R_\mathrm{s}$, where $R_\mathrm{s}$ is the (unique) radius where\footnote{We use $\log_{10}$ to denote the base-10 logarithm and $\log$ for the natural logarithm.} $\mathrm{d}\log_{10}\rho/\mathrm{d}\log_{10}{R} = -2$. N-body simulations predict a tight correlation between these two parameters \citep[e.g.][]{2016MNRAS.460.1214L}. In contrast, the pISO profile is characterised by a central $\rho \propto R^{0}$ core and an outer density slope of $\rho \propto R^{-2}$. Similarly to the NFW profile, the pISO model can also be characterised by two parameters. We choose an unconventional parametrization in analogy with the NFW parameters. The halo mass $M_\mathrm{200c}$ is defined in the same way as for the NFW profile, and we define a `concentration' as $c_\mathrm{pISO} = R_\mathrm{200c}/R_\mathrm{c}$, where $R_\mathrm{c}$ is the core radius where the density has fallen to half of the central asymptotic value. Unlike for $c_\mathrm{NFW}$, there is no strong prior for the relationship between $M_\mathrm{200c}$ and $c_\mathrm{pISO}$ \citep[but see][]{2004IAUS..220..377K}.

With these definitions the two halo models have very similar mathematical forms. $v_\mathrm{DM}$ can be expressed as
\begin{equation}
    v^2_\mathrm{DM}(R) = v^2_\mathrm{200c}\frac{R_\mathrm{200c}}{R}\frac{f_c(\frac{cR}{R_\mathrm{200c}})}{f_c(c)} \; ,
\end{equation}
where
\begin{align} 
    v^2_\mathrm{200c} &= \frac{GM_\mathrm{200c}}{R_{200c}} \; ,\\
    R_\mathrm{200c} &= \left(\frac{2GM_\mathrm{200c}}{200H^2_0}\right)^{\frac{1}{3}} \; ,
\end{align}
and $c$ is either $c_\mathrm{NFW}$ or $c_\mathrm{pISO}$ depending on the model. The $f_c$ function is defined for the NFW and pISO profiles as
\begin{equation}
    f_{c,\text{NFW}}(x) = \frac{\log(1+x)-x}{1+x}
\end{equation}
and
\begin{equation}
    f_{c,\text{pISO}}(x) = 1 - \frac{\arctan(x)}{x} \; ,
\end{equation}
respectively. This leaves $c_\mathrm{NFW}$ or $c_\mathrm{pISO}$, $M_\mathrm{200c}$, $\Upsilon_\mathrm{disc}$ and $\Upsilon_\mathrm{bulge}$ as the free parameters of our mass model that does not account for possible correlations in the rotation curve data.

To introduce the possibility of radial correlations in the rotation curves into our model, we use the parametric covariance matrix of \citet{2022RNAAS...6..233P}. The elements $K_{ij}$ of the matrix are given by the kernel function 
\begin{equation}
    \mathrm{K}_{ij} = k(R_i, R_j) + \sigma^2_{\mathrm{rot}, i}\delta_{ij} \; ,
\end{equation}
where
\begin{equation}
    k(R_i,R_j) = a_k^2\exp\left[-\frac{1}{2}\left(\frac{|R_i - R_j|}{s_k}\right)^2\right] \; , \label{eq:kernel}
\end{equation}
and $\delta_{ij}$ is the Kronecker delta function. $a_k$ and $s_k$ are the characteristic amplitude\footnote{The $A_k$ of \citet{2022RNAAS...6..233P} is equal to our $a_k^2$.} (e.g. in $\mathrm{km}\,\mathrm{s}^{-1}$) and scale (e.g. in $\mathrm{kpc}$) of the correlation. $a_k$ is the maximum correlation amplitude, for points separated by distances much smaller than $s_k$. The strength of the correlation smoothly tends to zero when the separation is much greater than $s_k$. The correlations, encoded by $k(R_i, R_j)$, are treated as separable from the uncertainties on the individual measurements, $\sigma_{\mathrm{rot},i}$.

This parametrization of the covariance in the data is empirical and can (and will) capture any of the sources of covariance outline in Sec.~\ref{sec:intro}. We caution that the true covariance is unlikely to have the exact form given by the kernel in Eq.~(\ref{eq:kernel}), which can lead to biases in model fitting. We explore the influence of the choice of kernel in Appendix~\ref{appendix:kernel}.

\subsection{Fitting method} \label{subsec:fitting}

We estimate the posterior probability distributions for our model parameters using Markov-Chain Monte Carlo (MCMC) sampling. For our models that neglect the possibility of correlations in the rotation curves we sample only the parameters $\theta_v = \left(\log_{10} M_\mathrm{200c},\log_{10} c, \log_{10}\Upsilon_\mathrm{disc},\log_{10}\Upsilon_\mathrm{bulge}\right)$ -- in galaxies without a bulge $\Upsilon_\mathrm{bulge}$ is omitted. We adopt log-uniform priors: $8\leq\log_{10}\left(M_\mathrm{200c}/\mathrm{M}_\odot\right)\leq 14$ and $0\leq\log_{10} c_\mathrm{NFW}\leq 3$ or $0\leq\log_{10} c_\mathrm{pISO}\leq 3$. For the mass-to-light ratios, we adopt log-normal priors peaking at $\log_{10}\left(\Upsilon_\mathrm{disc}/\mathrm{M}_\odot\,\mathrm{L}_\odot^{-1}\right)=-0.3$ and $\log_{10}\left(\Upsilon_\mathrm{bulge}/\mathrm{M}_\odot\,\mathrm{L}_\odot^{-1}\right)=-0.15$ with a standard deviation of $0.2\,\mathrm{dex}$ \citep[following][]{2016AJ....152..157L}. To focus on the influence of the treatment of radial correlations in rotation curves on the halo parameters, we fix the distances and inclinations of galaxies to the values tabulated in the SPARC compilation. The logarithmic likelihood for these models is given by \citep{2019A&A...626A..56P}:
\begin{equation}
    \log_{10} \mathcal{L}(\theta_v) = - \frac{1}{2} \sum_i \frac{\left[v_{\mathrm{rot},i} - v_\mathrm{model}(R_i|\theta_v)\right]^2}{\sigma^2_{\mathrm{rot},i}} \; .
\end{equation}
To fit the characteristic length and amplitude, we repeat the MCMC sampling with all six parameters: $\theta_k=\left(\log_{10} M_\mathrm{200c},\log_{10} c, \log_{10}\Upsilon_\mathrm{disc},\log_{10}\Upsilon_\mathrm{bulge},\log_{10} a_k,\log_{10} s_k\right)$. We keep the same priors as above for the parameters in common and impose log-uniform priors on $-2\leq\log_{10}\left(a_k/\mathrm{km}\,\mathrm{s}^{-1}\right)\leq 2.5$ and $-2\leq\log_{10}\left(s_k/\mathrm{kpc}\right)\leq 3$. Our choice of priors is discussed further in Appendix~\ref{appendix:priors}. The log-likelihood is then
\begin{equation}
    \log_{10} \mathcal{L}(\theta_k) = - \frac{1}{2}\mathrm{V}^\mathrm{T}_\mathrm{res} \mathrm{K}^{-1} \mathrm{V}_\mathrm{res} - \frac{1}{2}\log_{10} |\mathrm{K}| \; ,
\end{equation}
where $\mathrm{V}_\mathrm{res}$ is a vector with components $V_{\mathrm{res},i} = v_{\mathrm{rot},i}-v_\mathrm{model}(R_i|\theta_k)$. This is a GP regression problem.

\subsection{Configuration \& convergence of the MCMC sampler} \label{subsec:convergence}

We draw MCMC samples using \textsc{numpyro}'s \citep{2019arXiv191211554P} \textsc{nuts}\footnote{`No U-turn sampler', an adaptive variant of the Hamiltonian Monte-Carlo sampler.} sampler with \texttt{dense\_mass=True} and an acceptance probability of $0.9$; other configurable parameters are set to their default values. All MCMC sampling chains (3 in parallel for each model) were run for $4000$ steps with the first $1000$ discarded as burn-in. We visually inspected the MCMC chains to confirm that they converge to the likelihood peak before the end of the burn-in phase. We also tested running longer chains with longer burn-in phases, which led to no significant differences. We checked that our results are insensitive to reasonable variations in the number of radial sampling points used in constructing the GP model ($1000$ by default). For all galaxies in our sample, the sampling produces an average $\hat{R} = 1.03$, with $92.5$~per~cent satisfying $\hat{R} \leq 1.05$ and $66.9$~per~cent achieving $\hat{R} = 1.00$. The average effective sample size across all parameters is $N_\mathrm{eff} = 1831.66$.

\subsection{Example: DDO~154} \label{subsec:ddo154}

\begin{figure*}
    \centering
    \includegraphics[width=1\linewidth]{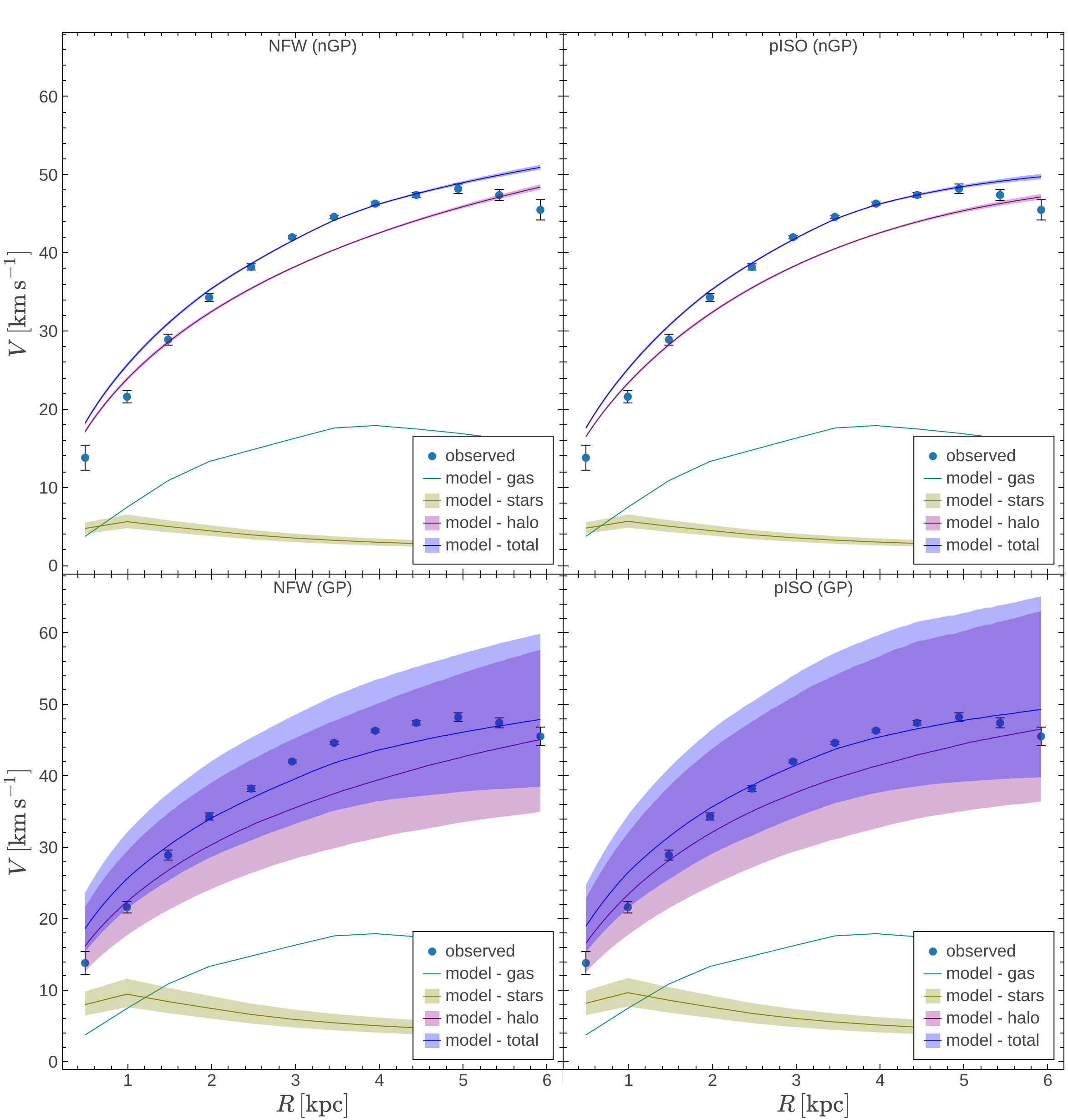}
    \caption{Mass models of DDO~154 for four cases. \textit{Upper left:} NFW halo without GP model for correlations in rotation curve. \textit{Upper right:} pISO halo without GP model. \textit{Lower left:} NFW halo with GP model. \textit{Lower right:} pISO halo with GP model. In all panels rotation curve measurements from the SPARC database are shown as points with error bars. The mass models show the gas (green line), stars (yellow line with shaded band) and halo (pink line with shaded band) components, along with the total model rotation curve (blue line with shaded band). The shaded bands show the $16^\mathrm{th}$-$84^\mathrm{th}$ percentile range of MCMC model evaluations at each radius. For both models, our view is that the GP case provides a more realistic description of models compatible with the data. Similar figures showing mass models for all $134$ galaxies in our sample are available as Supplementary Material.}
    \label{fig:DDO154-full}
\end{figure*}

Fig.~ \ref{fig:DDO154-full} shows the result of applying this fitting process to the rotation curve of DDO~154 -- one of the earliest instances of a `cored' dwarf galaxy \citep{1994ApJ...427L...1F,1994Natur.370..629M} -- as an illustrative example, for both the NFW and pISO halo model, with (GP) and without (nGP) the Gaussian process regression that allows for correlations in the rotation curve. In the nGP case the likely parameter values are strongly constrained by the few rotation curve points with the smallest uncertainties, resulting in an unrealistically small uncertainty on the best-fitting rotation curve and on the model parameters. Loosely speaking, the GP model responds by assuming a correlation between points such that a modest difference between the model and a few adjacent points with small uncertainties is penalized once instead of once per data point. This leads to a larger likely volume in the parameter space and correspondingly larger uncertainties on the best-fitting rotation curves.

\begin{table*}
    \caption{Parameter values for DDO~154 for the four mass models shown in Fig.~\ref{fig:DDO154-full}. Column descriptions: (1) the halo model; (2) whether the Gaussian process (GP) model for correlations in the rotation curve data is used or not (nGP); (3) halo mass; (4) $c$, the halo concentration (NFW model) or ratio of $R_\mathrm{200c}$ and core radius (pISO model); (5) stellar disc mass to light ratio; (6) amplitude of correlations in rotation curve (GP models only); (7) scale length for correlations in rotation curve (GP models only); (8) goodness of fit. All values are medians, with $16^\mathrm{th}$-$84^\mathrm{th}$ percentile intervals as uncertainties where applicable.}
    \centering
    \begin{tabular}{lccccccc}
        \hline
        Halo model & GP? & $\log_{10}(M_\mathrm{200c}/\mathrm{M}_\odot)$ & $\log_{10} c$ & $\log_{10}(\Upsilon_\mathrm{disc}/\mathrm{M}_\odot\,\mathrm{L}_\odot^{-1})$ & $\log_{10}(a_k/\mathrm{km}\,\mathrm{s}^{-1})$ & $\log_{10}(s_k/\mathrm{kpc})$ & $\chi^2_\mathrm{r}$\\
        \hline
        NFW & nGP &  $10.80 _{-0.06} ^{+0.06}$ & $0.74 _{-0.03} ^{+0.02}$ & $-0.83 _{-0.14} ^{+0.13}$  & -- & -- & 10.1\\[2ex]
        NFW & GP & $10.87 _{-0.77} ^{+1.09}$ & $0.65 _{-0.42}^{+0.34}$ & $-0.38_{-0.19}^{+0.18}$ & $1.17_{-0.41}^{+0.50}$ & $0.59_{-0.26}^{+0.26}$ & 1.2 \\[2ex]
        pISO & nGP & $10.10_{-0.03}^{+0.04}$ & $0.87_{-0.01}^{+0.01}$ & $-0.82_{-0.14}^{+0.13}$ & -- & -- & 6.7\\[2ex]
        pISO & GP & $10.26_{-0.52}^{+1.82}$ & $0.84_{-0.50}^{+0.18}$ & $-0.36_{-0.20}^{+0.17}$ & $1.14_{-0.41}^{+0.53}$ & $0.63_{-0.27}^{+0.38}$ & 1.3\\
        \hline
    \end{tabular}
    \label{tab:DDO154-params}
\end{table*}

The 1- and 2-dimensional marginalized posterior probability distributions for the model parameters for DDO~154 for both (GP and nGP) cases can be found in Appendix~\ref{appendix:corner}. The median values and $16^\mathrm{th}$-$84^\mathrm{th}$ percentile intervals are summarised in Table~\ref{tab:DDO154-params}. Although the most likely parameter values differ for parameters in common between the GP and nGP cases, they are consistent within the uncertainties (mostly due to the larger uncertainties on the GP parameter values).

\begin{figure*}
    \centering
    \includegraphics[width=\linewidth]{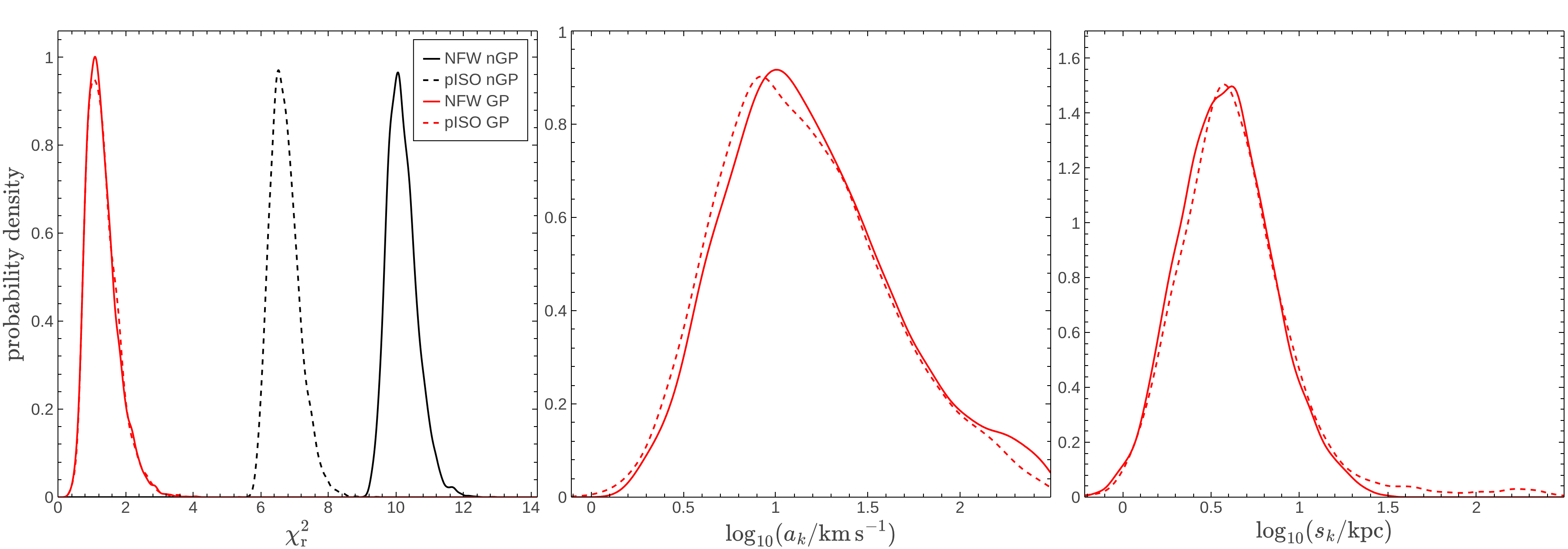}
    \caption{Goodness of fit and correlation amplitudes \& scale lengths for DDO~154. \textit{Left panel:} Reduced chi-squared $\chi^2_\mathrm{r}$ for each combination of halo model (NFW, solid lines; pISO, dashed lines) and Gaussian process model for radial correlations in the rotation curve (GP, red; nGP, black). In the nGP case there is a clear preference for the pISO halo model (lower typical $\chi^2_\mathrm{r}$), although neither model is a good fit ($\chi^2_\mathrm{r}\gg 1$). This preference vanishes when the GP model is used. \textit{Middle panel:} Marginalized posterior probability distribution for the amplitude of radial correlations, $a_k$, in the rotation curve. Lines are as in left panel. \textit{Right panel}: Marginalized posterior probability distribution for the scale length, $s_k$, of radial correlations in the rotation curve. Lines as in the left panel. The $a_k$ and $s_k$ distributions for the two models are nearly identical. All distributions are smoothed using Gaussian kernel density estimation (KDE).}
    \label{fig:DDO154-important}
\end{figure*}

An important difference between the nGP and GP models is the goodness of fit, which we quantify using the distribution of the reduced chi-squared statistic, $\chi^2_\mathrm{r}=\frac{1}{N_\mathrm{dof}}\left(\mathrm{V}_\mathrm{res}^\mathrm{T} \mathrm{K}^{-1} \mathrm{V}_\mathrm{res}\right)$, which we evaluate for each MCMC sample (see Appendix~\ref{app:goodness-of-fit} for a discussion of this choice of statistic for goodness of fit). $N_\mathrm{dof}$ is the number of degrees of freedom in the model. The left panel of Fig.~\ref{fig:DDO154-important} shows this distribution for the four models. Considering first the nGP models, the pISO halo provides a significantly better fit ($\chi^2_\mathrm{r}\sim 7$) than the NFW halo ($\chi^2_\mathrm{r}\sim 10$). This type of evidence has been used \citep[e.g.][see also \citealp{2023arXiv231020272M} for a different approach to model comparison]{2001AJ....122.2396D,2006ApJS..165..461K,2008MNRAS.383..297S,2008AJ....136.2648D,2015AJ....149..180O,2017MNRAS.466.1648K,2020ApJS..247...31L} to argue that cored halo models are statistically preferred over halo models with cusps, although in this example neither model achieves a good fit (i.e. $\chi^2_\mathrm{r}\sim 1$).

Turning next to the GP models, we find that both halo models result in a statistically good fit with a peak in the $\chi^2_\mathrm{r}$ distribution near $1$. This is not surprising since these models have considerably more freedom through the correlation parameters $a_k$ and $s_k$. However, the way that this is arises is important. For both halo models, the $a_k$ and $s_k$ parameters converge around maximum-likelihood values (see Appendix~\ref{appendix:corner}), and interestingly there is no strong covariance between these and $M_\mathrm{200c}$, $c$, or the mass-to-light ratios (although $a_k$ and $s_k$ are strongly covariant). Most interestingly, nearly exactly the same correlation strength and scale are preferred for both halo models, as shown in the centre and right panels of Fig.~\ref{fig:DDO154-important}. A slightly ($0.1\,\mathrm{dex}$, i.e. $\sim25$~per~cent) weaker correlation amplitude is preferred for the pISO than for the NFW halo model, but this is much less than the width of the distribution, and only about $1.5\,\mathrm{km}\,\mathrm{s}^{-1}$ in absolute terms. This slightly larger correlation amplitude could be responsible for making up the relative difference in goodness of fit in the nGP and GP cases between the two halo models, but we will see below that this is not typical across a larger sample of galaxies.

For the case of DDO~154 the conclusion is straightforward. If we accept that the rotation curve points could be correlated at around the level implied by the most likely $a_k$ and $s_k$ parameters, then there is no statistical preference for one halo model over the other. This possibility is made more compelling by the fact that both models achieve a good fit with nearly the same correlation amplitude and length scale, and our expectation that such correlations must be present (Sec.~\ref{sec:intro}), albeit with their strength unknown \textit{a priori}. We next turn to exploring the implications of allowing for such correlations by applying our modelling procedure to a larger sample drawn from the SPARC compilation.

\subsection{Sample selection} \label{subsec:sample}

After obtaining best-fitting parameters for the models defined in Sec.~\ref{subsec:massmodel} we pruned the initial SPARC sample of $175$ to those useful for further discussion of the influence of radial correlations in rotation curve data. We noticed that when the shape of the circular velocity curves allowed by a halo model (NFW or pISO) is incompatible with the rotation curve to be fit, the models including correlations (GP) respond by sampling unrealistically large values of $a_k$ and $s_k$. Qualitatively speaking, it is trying to achieve a formally good fit by penalising the shape mismatch only `once' (because all data are assumed to be extremely strongly correlated). In such cases a different, perhaps more flexible halo model might be needed to obtain a good fit with physically plausible values for $a_k$ and $s_k$. (Or, the rotation curve may not be a good tracer of the mass distribution, for example because the galaxy is out of equilibrium.) We therefore restrict our sample of galaxies to those where a plausible fit of at least one of the NFW and pISO halo models is possible.

\begin{figure}
    \centering
    \includegraphics[width=\linewidth]{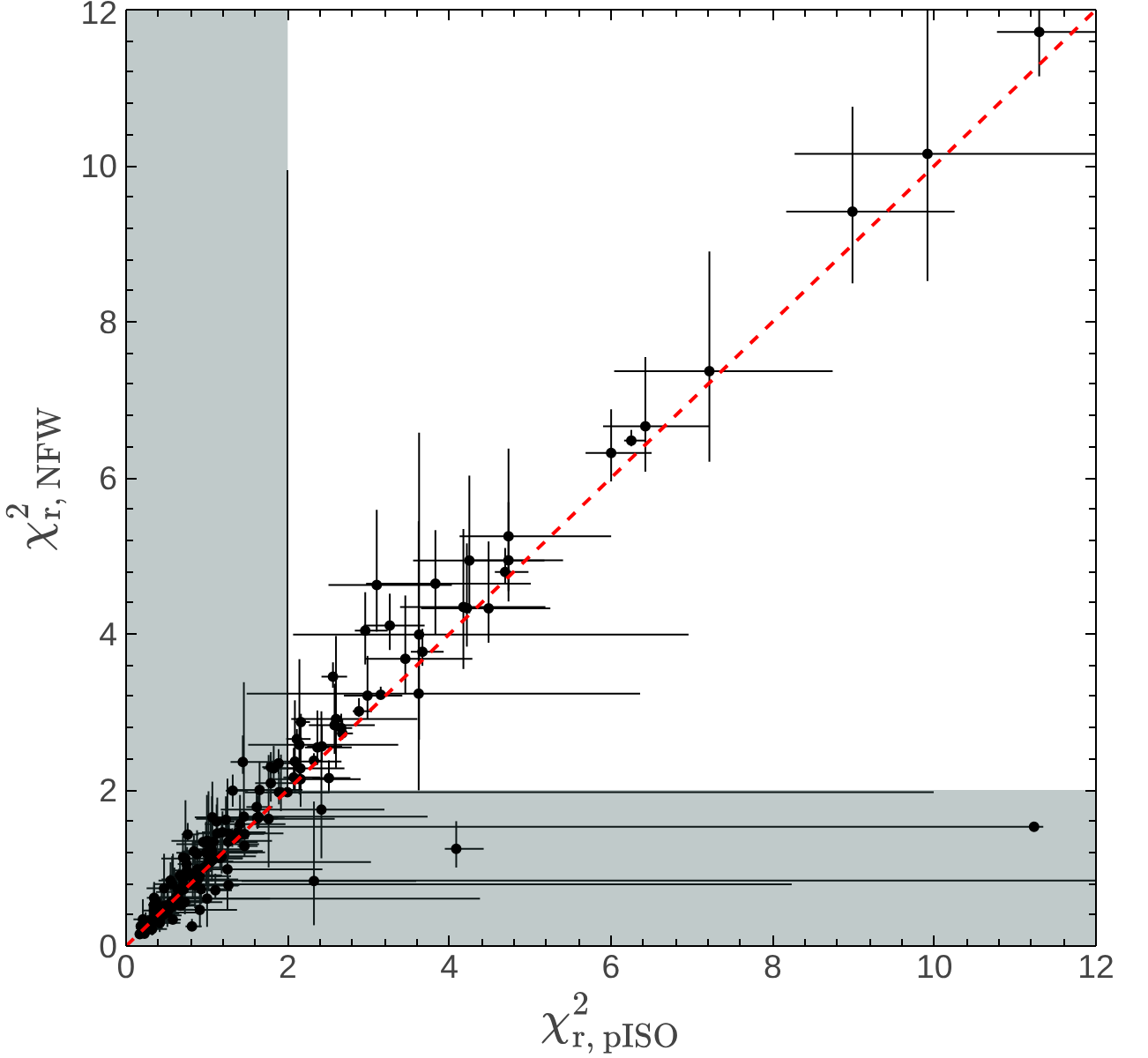}
    \caption{Comparison of the $\chi^2_\mathrm{r}$ values for the pISO (abscissa) and NFW (ordinate) halo models when constant uncertainties of $0.05\max(v_\mathrm{rot})$ are assumed on the rotation curve measurements. The points mark the median of the distribution across the MCMC samples with the $16^\mathrm{th}$-$84^\mathrm{th}$ percentile intervals shown by the error bars. The dashed red line shows the $1$:$1$ relation. A majority of the points lie above this line, indicating a preference for the pISO halo model. We select all galaxies whose rotation curve shapes can be well captured by at least one of the two halo models -- where at least one achieves a median $\chi^2_\mathrm{r} < 2$, shown as the shaded region.}
    \label{fig:sample-selection}
\end{figure}

The goodness of fit (parametrised by $\chi^2_\mathrm{r}$) for our models without allowing for radial correlations (nGP) are in principle a good metric to identify galaxies where a good fit cannot be achieved by either/both models. However, many galaxies whose rotation curve shapes could be described well by one or both halo models have small uncertainties in  $v_\mathrm{rot}$ that result in much larger $\chi^2_\mathrm{r}$ values than for galaxies with similar rotation curve shapes but larger uncertainties. DDO~154 is one example of such a galaxy. Therefore, for the purpose of selecting our galaxy sample only, we adopt a constant absolute uncertainty of $5$~per~cent of the galaxy's rotation curve maximum for each $v_\mathrm{rot}$ measurement. These are typically somewhat larger than the nominal uncertainties recorded in the SPARC compilation. For each galaxy, the median value of the $\chi^2_\mathrm{r}$ posterior distributions of the nGP models (with these constant uncertainties) for both halo models are plotted against each other in Fig.~\ref{fig:sample-selection}. The $16^{\text{th}}$ and $84^{\text{th}}$ percentiles of the distributions are shown with error bars. Any galaxy with a median $\chi^2_\mathrm{r}$ value less than $2$ -- indicating that the relevant halo model can provide a reasonable match for the shape of the rotation curve -- for either halo model is accepted into our sample.

After this initial selection, we also remove any remaining galaxies with less than $10$ velocity measurements. As the GP models have up to six fitting parameters, this would leave a maximum of three degrees of freedom in the model. We found that keeping the relative change in the number of degrees of freedom between the nGP and GP models small -- by keeping the total number of degrees of freedom relatively large -- simplifies interpreting the resulting best-fitting models.

This selection process removes $41$ galaxies, $28$ of which are removed for having $<10$ velocity measurements after the initial selection. The SPARC compilation includes a `quality flag' $Q$ for each galaxy, providing a heuristic measure of the quality of the H\,\textsc{i} or H$\alpha$ data, and the probability that a rotation curve will be a good tracer of the circular velocity. For example, very asymmetric galaxies, galaxies with strong non-circular motions and galaxies with gas spatially offset from their stars are penalized in $Q$. Our cuts remove $75$~per~cent of the galaxies flagged $Q=3$ by \citet{2016AJ....152..157L} from our sample (we do not explicitly use this quality flag in our selection). $45$~per~cent of $Q=2$ galaxies and only $3$~per~cent of $Q=1$ galaxies are also removed, providing some reassurance that the discarded galaxies are generally less suitable for mass modelling overall. Our final sample comprises $134$ galaxies.

\section{Results} \label{sec:results}

We next proceed to apply the same analysis as for DDO~154 above (Sec.~\ref{subsec:ddo154}) to our selection from the SPARC sample. Details of all models are included as Supplementary Material. Our overall impression is that the uncertainties in the GP mass models are more representative of what can plausibly be inferred from the data (i.e. they are larger) than those in the nGP models. We begin by examining the quality of the fits, parametrised by the $\chi^2_\mathrm{r}$ statistic (Sec.~\ref{subsec:chi2}), then consider the posterior probability distributions for the two additional parameters ($a_k$ and $s_k$) in the GP case (Sec.~\ref{subsec:aksk}).

\subsection{Goodness of fit} \label{subsec:chi2}

\begin{figure*}
    \centering
    \includegraphics[width=\linewidth]{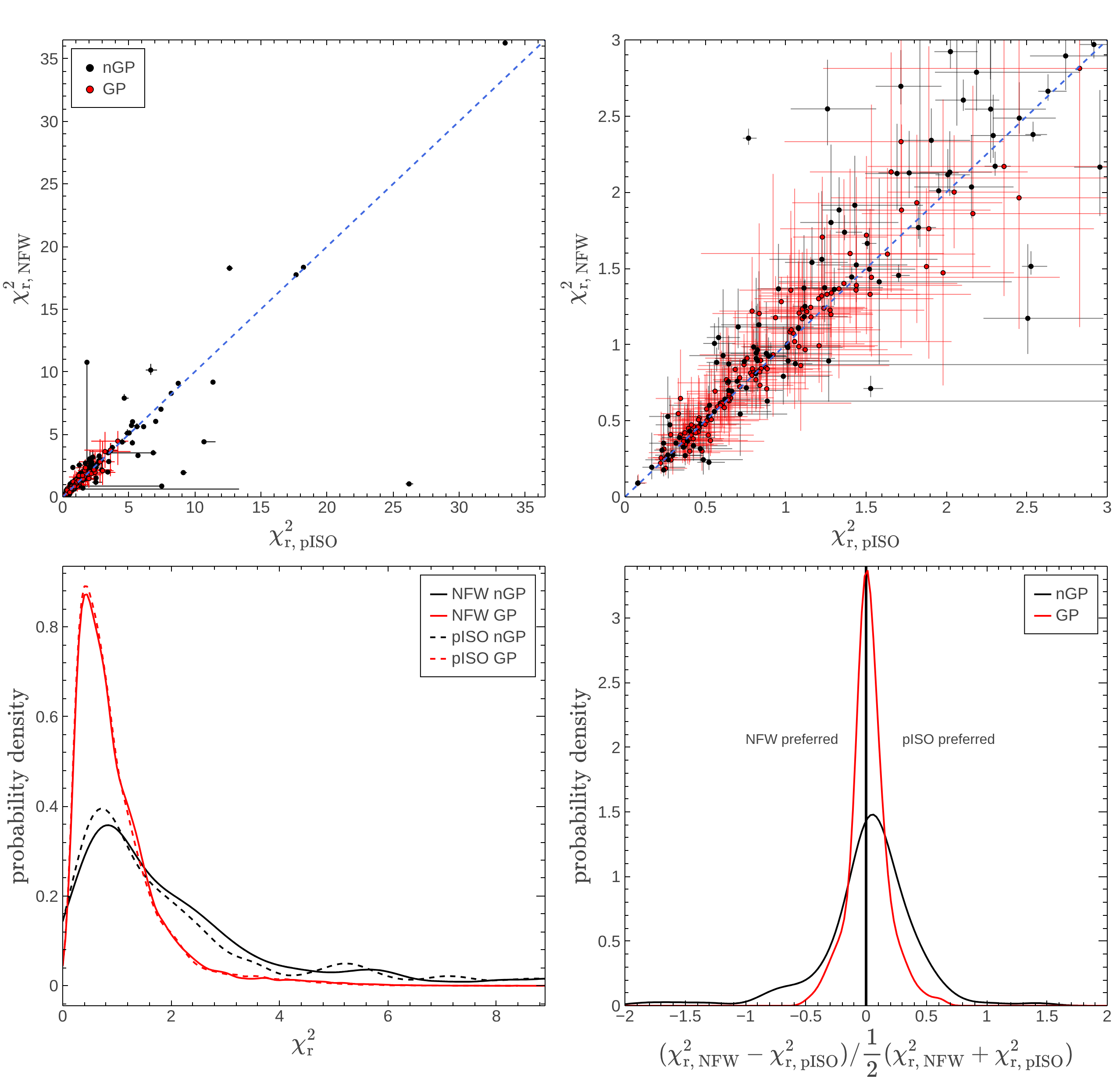}
    \caption{\textit{Upper left:} Comparison of the $\chi^2_\mathrm{r}$ values for the pISO (abscissa) and NFW (ordinate) halo models with (red) and without (black) the GP model for radial correlations in the rotation curves. The points mark the median of the distribution across the MCMC samples with the $16^\mathrm{th}-84^\mathrm{th}$ percentile intervals shown by the error bars. The dashed red line shows the $1$:$1$ relation. The values are tabulated in Table~\ref{tab:all-chi2}. \textit{Upper right:} as upper left but cropped to the $\chi^2_\mathrm{r} < 3$ region. In the nGP case most points lie above the $1$:$1$ line indicating a preference for the pISO model, while in the GP case the points are evenly distributed around it, indicating no preference for either model. \textit{Lower left:} distribution of $\chi^2_\mathrm{r}$ summed across all sampled galaxies, smoothed with Gaussian KDE. The same slight preference for the pISO model in the nGP case as noted in the description of the upper left panel is visible here as a slightly sharper peak near $\chi^2_\mathrm{r}=1$. In the GP case the distributions for the two halo models are almost identical and much more sharply peaked near $\chi^2_\mathrm{r}=1$ than their nGP counterparts. \textit{Lower right:} The distribution of the fractional difference between the median $\chi^2_\mathrm{r}$ values for each galaxy. The distribution for the nGP case is slightly offset towards $\chi^2_\mathrm{r,pISO}<\chi^2_\mathrm{r,NFW}$, again indicating a slight preference for the pISO halo model. This confirms that the preference seen in the other panels is not driven only by a very strong preference for the pISO model in a small number of galaxies. In the GP case the distribution is centred about $0$, indicating no preference for one halo model over the other.}
    \label{fig:main-results}
\end{figure*}

Fig.~\ref{fig:main-results} shows the median $\chi^2_\mathrm{r}$ value for the pISO halo model compared to that for the NFW halo model for both the nGP and GP cases (upper left panel). The upper right panel provides a closer look at the $\chi^2_\mathrm{r} < 3$ region. Since we have selected only galaxies where at least one model can provide a good description of the shape of the rotation curve (see Sec.~\ref{subsec:sample}), the cases with large $\chi^2_\mathrm{r}$ for both models necessarily come from galaxies where there is a good match to the overall shape but the uncertainties on individual points are very small. These only occur in the nGP models. The GP model responds to this situation by increasing $a_k$ and/or $s_k$ until a statistically good fit is achieved -- $\chi^2_\mathrm{r}$ is never more than $\sim 2$ for this selection of galaxies. We recall that our sample includes galaxies where at least one of the two models can broadly match the rotation curve shape, the other model may fail catastrophically. The most extreme example of this is UGC~02885 with $\chi^2_\mathrm{r,\,pISO}\sim 26$ but $\chi^2_\mathrm{r,\,NFW}\sim 1$ (in the nGP case).

Statistical preference for one model over the other at the galaxy population level can be qualitatively assessed by comparing the $\chi^2_\mathrm{r}$ distributions summed over the entire sample. This is shown in the lower left panel of Fig.~\ref{fig:main-results}. In the nGP case the distribution for the pISO model is slightly more strongly peaked around $\chi^2_\mathrm{r}=1$ than that for the NFW model, indicating a very slight preference for the pISO over the NFW halo model (on average, $\chi^2_\mathrm{r,\,pISO} < \chi^2_\mathrm{r,\,NFW}$). The lower right panel of the same figure shows that this is not due to a few outliers distorting the distribution but is also true on a galaxy-by-galaxy basis. The panel shows that the fractional difference:
\begin{equation}
    \frac{\chi^2_\mathrm{r,\,NFW} - \chi^2_\mathrm{r,\,pISO}}{\frac{1}{2}\left(\chi^2_\mathrm{r,\,NFW} + \chi^2_\mathrm{r,\,pISO}\right)}
\end{equation}
evaluated galaxy-by-galaxy is slightly shifted towards a preference for the pISO halo model ($88$/$134$ galaxies; $65.7$~per~cent). That this preference is so weak came as a surprise in light of the claims of a clear preference for the pISO model of the NFW model in the literature \citep[e.g.][]{2008AJ....136.2648D,2020ApJS..247...31L} -- we will return to this point in Sec.~\ref{sec:pISO-preference} below.

In the GP models, the slight preference vanishes: the pISO and NFW halo models fare equally well. The overall $\chi^2_\mathrm{r}$ distributions are essentially indistinguishable and peak near unity, and the fractional difference distribution is very nearly symmetric around $0$. For the NFW halo model, $75$ galaxies ($56$~per~cent) have improved fits compared to the nGP case. Similarly, $76$ galaxies ($57$~per~cent) are improved in the GP case for the pISO halo model. The fits for the remaining galaxies do not get worse -- these already had good fits in the nGP case ($\chi^2_\mathrm{r} \approx 1$).

A list of all galaxies in the sample and their median $\chi^2_\mathrm{r}$ for all four cases are given in Appendix~\ref{appendix:table}. We also include a complete set of figures including mass models (as in Fig.~\ref{fig:DDO154-full}), corner plots (as for DDO~154 in Appendix~\ref{appendix:corner}), and $\chi^2_\mathrm{r}$ distributions (as in Fig.~\ref{fig:DDO154-important}) for all $134$ galaxies in our sample as Supplementary Material.

\subsection{Correlation amplitudes and scales} \label{subsec:aksk}

\begin{figure}
    \centering
    \includegraphics[width=\linewidth]{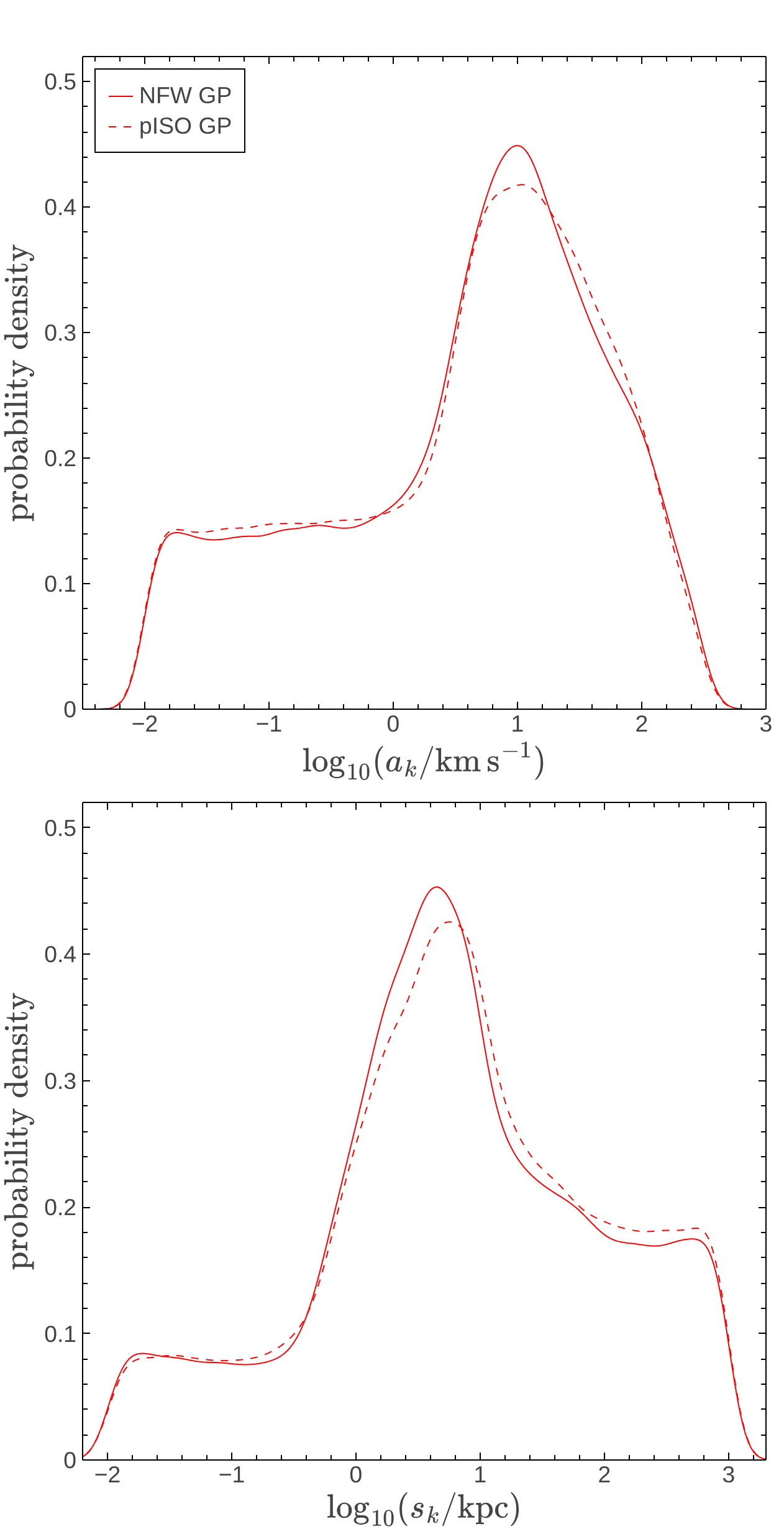}
    \caption{Posterior probability distributions for $a_k$ (upper panel) and $s_k$ (lower panel) summed across all $134$ galaxies in our sample. Distributions are smoothed using Gaussian KDE. Both halo models show very similar distributions for correlation amplitude and scale length; the pISO model has a preference for very slightly stronger and further-reaching correlations.}
    \label{fig:ak-sk}
\end{figure}

Fig.~\ref{fig:ak-sk} shows the posterior distributions of the GP models' characteristic correlation amplitude $a_k$ (upper panel) and scale $s_k$ (lower panel) for both halo models summed across all galaxies in our sample. The overall distributions are very similar. Interestingly there is a very slight preference for stronger correlations in the pISO model than in the NFW model. We might have expected that the disappearance in the GP models of the slight preference for the pISO over the NFW model found in the nGP models (Sec.~\ref{subsec:chi2}) was due to fitting slightly stronger correlations for the NFW model to compensate for the poorer fits in the nGP models -- this is not the case. The equal preference for both halo models in the GP case therefore seems to be a general consequence of modelling the correlations, rather than a systematic effect caused by the NFW halo GP model preferring stronger correlations to achieve better fits.

The GP models settle on preferred values for $a_k$ and $s_k$ that seem realistic (mostly $a_k\lesssim 30\,\mathrm{km}\,\mathrm{s}^{-1}$ and $s_k\lesssim 10\,\mathrm{kpc}$), although the MCMC chains for some galaxies (usually ones with especially small uncertainties on the rotation curve) do sample much larger values. Only one galaxy has a peak (most likely) value for $a_k$ above $56\,\mathrm{km}\,\mathrm{s}^{-1}$. We recall that we have selected galaxies where at least one of the halo models can reproduce the shape of the rotation curve. Attempting to fit the GP models to SPARC galaxies excluded from our sample often leads to artificially good fits with unphysically large values for $a_k$ and $s_k$ and very broad posterior probability distributions for the dark halo parameters.

\begin{figure*}
    \centering
    \includegraphics[width=1\linewidth]{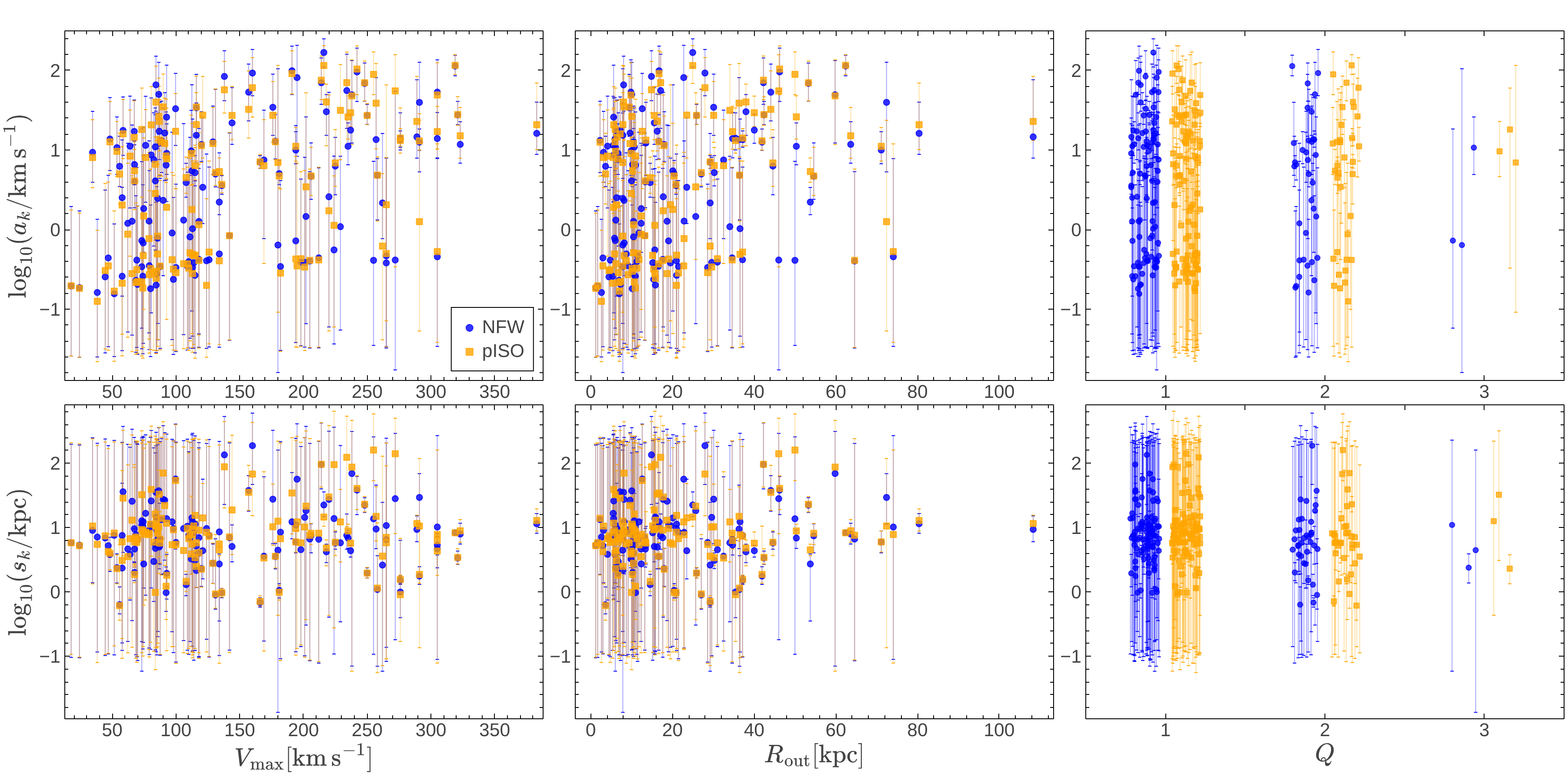}
    \caption{Correlation amplitude ($a_k$, top row) and scale ($s_k$, bottom row) dependence on sampling and resolution indicators across all 134 galaxies in our sample as a function of maximum rotation velocity ($V_\mathrm{max}$, left column), the radius of the outermost rotation curve measurement ($R_\mathrm{out}$, middle column) and SPARC quality flag ($Q$, right column; a small amount of horizontal scatter has been added for clarity). Points mark the $50^\mathrm{th}$ percentile of the posterior probability distribution, with $16^\mathrm{th}$-$84^\mathrm{th}$ percentile interval shown with error bars. There are no strong trends with these sampling metrics.}
    \label{fig:correlation-vmax-rout-q}
\end{figure*}

Fig.~\ref{fig:correlation-vmax-rout-q} shows the correlation amplitude and scale as a function of maximum rotation velocity $V_\mathrm{max}$, radius of the outermost measurement $R_\mathrm{out}$, and SPARC data quality flag $Q$. There are no strong trends with any of these metrics, suggesting that the inference is not dominated by galaxy properties or data quality. There is weak structure in $a_k$ with $V_\mathrm{max}$ and $R_\mathrm{out}$. This is not unexpected: a galaxy with higher $V_\mathrm{max}$ may span a larger range in velocity across the rotation curve, providing a physical motivation for the larger correlation amplitudes inferred for some galaxies. Similarly, galaxies with larger $R_\mathrm{out}$ encapsulate a larger disk, where multiple or larger perturbations and other dynamical effects may contribute to increased inferred size of the correlation amplitudes. However, the very weak trends and large scatter in all panels of Fig.~\ref{fig:correlation-vmax-rout-q} imply that such effects are not the main drivers of the inferred values for $a_k$ and $s_k$.

\section{Discussion} \label{sec:discussion}

\subsection{The preference for the pISO halo} \label{sec:pISO-preference}

Our finding that there is no strong statistical preference for the cored pISO halo model over the cuspy NFW halo model in the nGP case was unexpected (Sec.~\ref{subsec:chi2}). \citet{2020ApJS..247...31L} previously fit mass models to all SPARC galaxies and reported $\chi^2_\mathrm{r}$ values. This provides a helpful touchstone for comparison since that study used the same velocity measurements and baryonic mass profiles (from SPARC) and the same dark halo models (pISO and NFW, amongst others), therefore differing only in the model fitting methodology. The models are similarly parametrized, with their $v_\mathrm{200c}$ taking the place of our $M_\mathrm{200c}$ (we do not expect this to introduce any appreciable bias), and exact analogues of our $c$, $\Upsilon_\mathrm{bulge}$ and $\Upsilon_\mathrm{disc}$. In contrast to our approach, distance and inclination were also included as free parameters, with Gaussian priors centred on the values tabulated in the SPARC compilation, using the uncertainty as the standard deviation. Their priors\footnote{They also consider cases where a prior is imposed on the joint distribution of $V_{200}$ and $C_{200}$ (their `$\Lambda$CDM' prior), but we compare with their `flat' prior case that is analogous to our choice of priors.} on $\log_{10}\left(\Upsilon_\mathrm{disc}/\mathrm{M}_\odot\,\mathrm{L}_\odot^{-1}\right)$ and $\log_{10}\left(\Upsilon_\mathrm{bulge}/\mathrm{M}_\odot\,\mathrm{L}_\odot^{-1}\right)$ differ from ours: they also adopt normal distributions centred on $-0.3$ for the disc and $-0.15$ for the bulge \citep[following][]{2016PhRvL.117t1101M,2017ApJ...836..152L}, but a narrower width of $0.1\,\mathrm{dex}$ (we used $0.2\,\mathrm{dex}$).

\begin{figure}
    \centering
    \includegraphics[width=\linewidth]{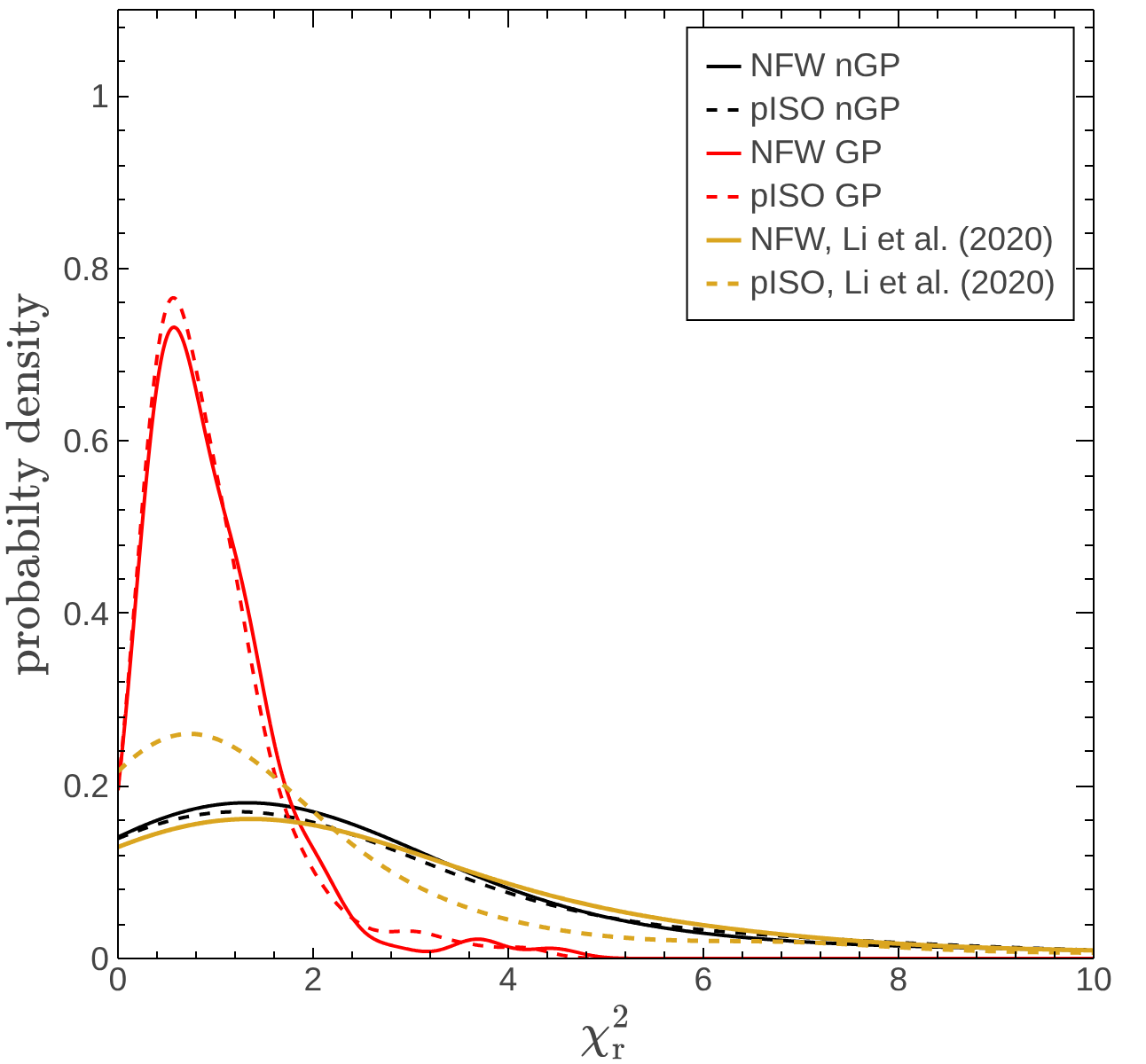}
    \caption{The distribution of the median $\chi^2_\mathrm{r}$ values for galaxies in our sample for the NFW (solid lines) and pISO halo model (dashed lines) when radial correlations in the rotation curves are modelled (GP; red) or not (nGP; black). This is similar to the lower left panel of Fig.~\ref{fig:main-results} but plotting the distribution of median values instead of the summed distribution allows us to compare with distributions of the $\chi^2_\mathrm{r}$ values reported by \citet{2020ApJS..247...31L} -- these are shown with the yellow lines. We use only the $134$ galaxies from our sample in constructing all distributions, and all distributions are Gaussian KDE smoothed. \citet{2020ApJS..247...31L} found a stronger preference for the pISO over the NFW halo model than we do, see Sec.~\ref{sec:pISO-preference} for a discussion.}
    \label{fig:freedoms}
\end{figure}

We show the distribution of the median $\chi^2_\mathrm{r}$ value of galaxies in our sample for both halo models and the GP and nGP cases in Fig.~\ref{fig:freedoms}, and compare with the same distribution as tabulated in \citet{2020ApJS..247...31L}. Their fits are directly comparable with our nGP fits, with only the differences in methodology noted above. Our distribution of $\chi^2_\mathrm{r}$ across the galaxy sample agrees very closely with theirs for the NFW halo model, but they find systematically lower $\chi^2_\mathrm{r}$ values for the pISO halo model. The only possible sources for this difference are the added freedom in distance and inclination (effectively allowing independent multiplicative scaling of the radial and velocity axes of the rotation curve and its components) and the tighter priors on the mass-to-light ratio parameters. Re-running our model fits with either modified priors or extended to include distance and inclination as free parameters with similar priors as in \citet{2020ApJS..247...31L} reveals that the latter is the primary driver of their preference for the pISO over the NFW model where we find none\footnote{Re-running our nGP model fits to be as similar as possible to the NFW-Flat and pISO-Flat models of \citet{2020ApJS..247...31L} -- i.e. including distance and inclination as free parameters and matching all priors -- we find that we qualitatively reproduce their stronger preference for the pISO over the NFW halo model. If we use exactly the same model in the GP case, both models achieve a good fit and there is no preference for one model over the other. We treat distance and inclination as fixed parameters throughout our analysis to focus attention on the influence of the model for radial correlations in the rotation curves.}. This highlights the sensitivity of such model comparisons to what could reasonably be called minor changes to the modelling approach.

\subsection{The halo mass-concentration relation} \label{subsec:mcr}

\begin{figure}
    \centering
    \includegraphics[width=\linewidth]{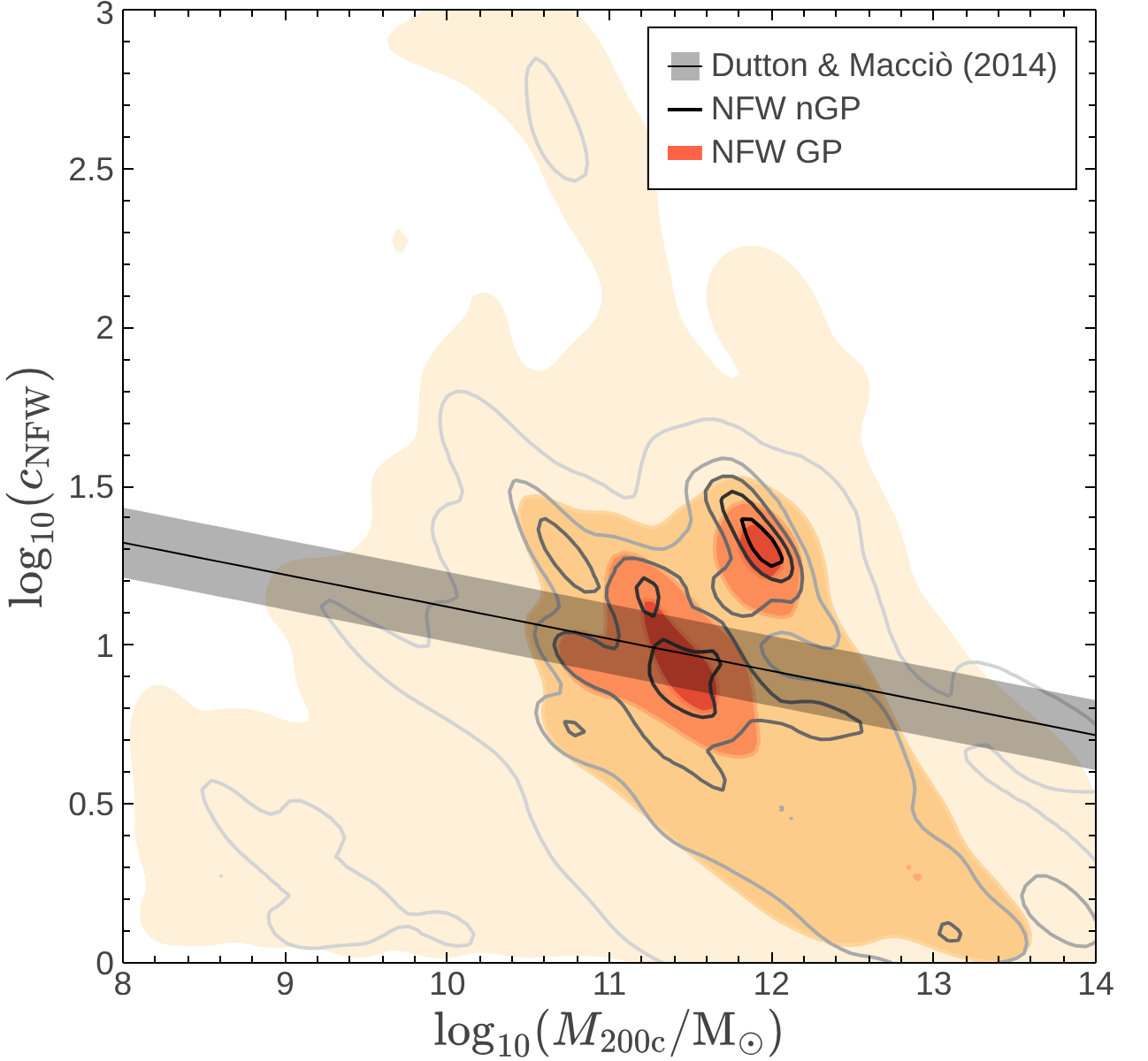}
    \caption{Two-dimensional marginalized posterior probability distributions for the $c_\mathrm{NFW}$ and $M_\mathrm{200c}$ parameters of the NFW halo model, summed across all galaxies in our sample. The nGP and GP cases are shown with open and filled linearly-spaced contours, respectively. The $M_\mathrm{200c}-c_\mathrm{NFW}$ relation of \citet{2014MNRAS.441.3359D} and its scatter are shown with the black line and shaded region. The probability density is slightly more concentrated around the $M_\mathrm{200c}-c_\mathrm{NFW}$ relation in the GP case, but several galaxies in the sample are still incompatible with the relation in both cases.}
    \label{fig:m200-c}
\end{figure}

We recall that we did not impose a prior on the mass-concentration relation for the NFW model (Sec.~\ref{subsec:fitting}) to enable a fairer comparison with the pISO halo model. However, we can check whether our posterior probability distributions are consistent with the expectation for $\Lambda$CDM haloes. The 2-dimensional marginalized posterior probability distribution for $M_\mathrm{200c}$ and $c_\mathrm{NFW}$ summed over our entire sample is shown in Fig.~\ref{fig:m200-c}. The distribution keeps the same overall shape in both the nGP and GP cases, but more weight shifts onto the peak on the \citet{2014MNRAS.441.3359D} relation near $M_\mathrm{200c}=10^{11.5}\,\mathrm{M}_\odot$ in the GP case. Allowing for possible correlations in the rotation curve data therefore seems to lead to a greater preference for the kinds of dark haloes predicted by $\Lambda$CDM N-body simulations. There are, however, still many samples well off the relation in both cases. The overall shape of the distribution in the nGP is quite similar to that found for the NFW model (flat priors) case by \citet[][see their fig.~3]{2020ApJS..247...31L}.

\section{Conclusions} \label{sec:conclusions}

The points in measured rotation curves are necessarily correlated with each other. These correlations arise for reasons including: (i) the definition of the circular velocity as an integral quantity; (ii) perturbative processes affecting galaxies; (iii) beam smearing; and likely other sources that are difficult to model. The amplitudes and length scales associated with such correlations are difficult to determine \textit{a priori} (Sec.~\ref{sec:intro}). We allow for correlations from any source but described by a parametric covariance matrix while constructing mass models to fit the rotation curves of galaxies from the SPARC compilation (Sec.~\ref{sec:method}), following the data-driven approach of \citet{2022RNAAS...6..233P}. In our view the resulting mass models have uncertainties more representative of the constraining power of the data than when correlations are neglected. The correlation amplitudes and scale lengths that our models prefer for SPARC galaxies are physically plausible, on the order of $20\,\mathrm{km}\,\mathrm{s}^{-1}$ and $5\,\mathrm{kpc}$, respectively (Sec.~\ref{subsec:aksk}). While correlations must be present at some level -- and our models will capture this -- it is somewhat indiscriminate: other effects, such as underestimated uncertainties on the rotation curve measurement, could be absorbed into inflated estimates of the correlation amplitude and scale length. We therefore consider our constraints on these parameters as upper limits.

Allowing for correlations erases the statistical preference for cored dark halo models (e.g. pISO) over ones with a central density cusp (e.g. NFW) that we find in our models that do not account for correlations and also reported in other work \citep[e.g.][]{2001AJ....122.2396D,2006ApJS..165..461K,2008MNRAS.383..297S,2008AJ....136.2648D,2015AJ....149..180O,2017MNRAS.466.1648K,2020ApJS..247...31L}. This also results a statistically good fit ($\chi^2_\mathrm{r}\approx 1$) for both models in many galaxies where neither model achieves a good fit ($\chi^2_\mathrm{r}\gg 1$) when correlations are ignored (Sec.~\ref{subsec:chi2}). This would be unsurprising if stronger correlations were needed to achieve an acceptable fit for the previously less favoured NFW model than for the pISO model, but this does not seem to be the case -- if anything, the pISO model prefers (slightly) stronger correlations.

The weakened preference for one model over the other when correlations are allowed for implies weakened constraints on the central density profiles of haloes where the NFW and pISO models differ most. In more flexible models \citep[][etc.]{1965TrAlm...5...87E,2014MNRAS.437..415D,2017MNRAS.468.1005D} the constraint on the parameter(s) controlling the central density slope will be weaker when correlations are allowed for, much like the constraints on the mass and concentration parameters explored in this work.

Comparison with the mass models of \citet{2020ApJS..247...31L} reveals that the statistical preference for for one model over another (e.g. the pISO over the NFW) can be very sensitive to the choice of priors (Sec.~\ref{sec:pISO-preference}). Our mass models that do not account for correlations in the rotation curves are very similar to theirs, except that we treat the distances and inclinations of galaxies as fixed to the values recorded in the SPARC tables whereas they allow some freedom within the recorded uncertainties. This apparently small difference is enough to change the preference for the pISO model over the NFW from very slight (in our models) to much stronger (in theirs). Allowing for radial correlations in the rotation curves results in no preference for one model over the other for any choice of these priors that we explored. This implies that careful handling of correlations in rotation curve data are essential to correctly interpret them, including when evaluating the evidence for dark matter cusps or cores.

Allowing for correlations also seems to bring the haloes inferred for many of the galaxies that we modelled closer to the locus in halo mass and concentration expected from cosmological N-body simulations in the $\Lambda$CDM cosmogony (Sec.~\ref{subsec:mcr}). Many galaxies in our sample still fall well off this locus however, and we also omit $41$ galaxies from the SPARC compilation that neither the NFW nor the pISO model can fit well even when correlations in the rotation curves are allowed for (Sec.~\ref{subsec:sample}). Other interpretations must be sought for these galaxies, and those may feed back into the interpretation of correlations in rotation curves of all galaxies.

Finally, we remark that the form assumed for the correlations in the covariance matrix used when constructing the mass models is amongst the simplest possible -- described by a single constant amplitude and scale length for each galaxy, with a squared-exponential kernel \citep[see also][for a discussion of other kernel choices -- we also find that this choice usually has little influence on the parameter inference]{2023ARA&A..61..329A}. Any correlations between velocity measurements at different radii are certainly not constant across the discs of galaxies, so our analysis should be taken as an assessment of the overall influence that accounting for this feature of the data could have on mass models of galaxies. We conclude that they are plausibly as influential as several other sources of systematic uncertainty in mass models and bear further investigation in future work.

\section*{Software}

The following software packages were used in this work:
\textsc{arviz} \citep{2019JOSS....4.1143K}, 
\textsc{astropy} \citep{2022ApJ...935..167A}, 
\textsc{bokeh}\footnote{docs.bokeh.org/en/latest/}, 
\textsc{contourpy}\footnote{github.com/contourpy/}, 
\textsc{corner} \citep{2016JOSS....1...24F}, 
\textsc{jax}\footnote{github.com/jax-ml/jax/}, 
\textsc{numpy} \citep{2020Natur.585..357H}, 
\textsc{numpyro} \citep{2019arXiv191211554P}, 
\textsc{scipy}\footnote{docs.scipy.org/doc/}, 
\textsc{tingp} \citep{2024zndo..10463641F}.

\section*{Acknowledgements}

HC was supported through the Air Cadet Development Scheme (ref: ACDS-26-005) and a Royal Astronomical Society summer studentship. KAO, DD \& KEH acknowledge support by the Royal Society through a Dorothy Hodgkin Fellowhsip (DHF/R1/231105) held by KAO. KAO also acknowledges support by the European Research Council (ERC) through an Advanced Investigator grant to C.~S.~Frenk, DMIDAS (GA 786910), and STFC through grant ST/T000244/1. This research has made use of NASA's Astrophysics Data System. 

For the purpose of open access, the author has applied a Creative Commons Attribution (CC BY) licence to any Author Accepted Manuscript version arising from this submission.

\section*{Data Availability}
 
The SPARC data used in this work \citep{2016AJ....152..157L} are publicly available from the Strasbourg astronomical Data Center\footnote{https://cds.unistra.fr} (CDS) with DOI \texttt{10.26093/cds/vizier.51520157}. The data from \citet{2020ApJS..247...31L} are available from the CDS with DOI \texttt{10.26093/cds/vizier.22470031}. The fitting routines of \citet{2022RNAAS...6..233P} are available at \url{https://lposti.github.io/MLPages/gaussian_processes/2022/11/02/gp_rotcurves.html}; our modified version of these is included with this article as Supplementary Material.

\bibliographystyle{mnras}
\bibliography{references}

\section{Supporting information}

We include as supplementary materials:
\begin{itemize}
\item A document containing Appendix~E collecting figures similar to Figs.~\ref{fig:DDO154-full}, \ref{fig:NFW-corner-plot} \& \ref{fig:pISO-corner-plot} for all $134$ galaxies in our sample.
\item A code file containing our fitting routines adapted from those accompanying \citet{2022RNAAS...6..233P}.
\end{itemize}

\appendix

\section{Goodness of fit statistic} \label{app:goodness-of-fit}

Throughout our analysis we use the distribution of the $\chi^2_\mathrm{r}$ statistic evaluated across MCMC samples as a measure of goodness of fit for our models. For the GP models it is not obvious \textit{a priori} that this is a good statistic because the covariance matrix $\mathrm{K}$ used to evaluate it depends on the model parameters $a_k$ and $s_k$. We therefore investigate an alternative statistic to quantify the goodness of fit based on whitened residuals. We use the Cholesky transformation of $\mathrm{K}$ to obtain a whitening transformation $\mathrm{K}^{-1/2}$, apply this to the model residuals as $\mathrm{K}^{-1/2}\mathrm{V}_\mathrm{res}$ for each MCMC sample, and assess the resulting (flattened) distribution of whitened residuals. This should follow a Gaussian distribution with a standard deviation of $1$ for a good model fit. The distributions for the GP models fit to some galaxies closely resemble such Gaussian distributions, while others differ in width or shape (e.g. skew or kurtosis), with extreme cases having multiple widely-spaced peaks, reflecting the range of goodness of fit of the models for different galaxies -- broadly this gives a similar impression to the distributions of the $\chi^2_\mathrm{r}$ statistic.

\begin{figure*}
  \centering
  \includegraphics[width=\linewidth]{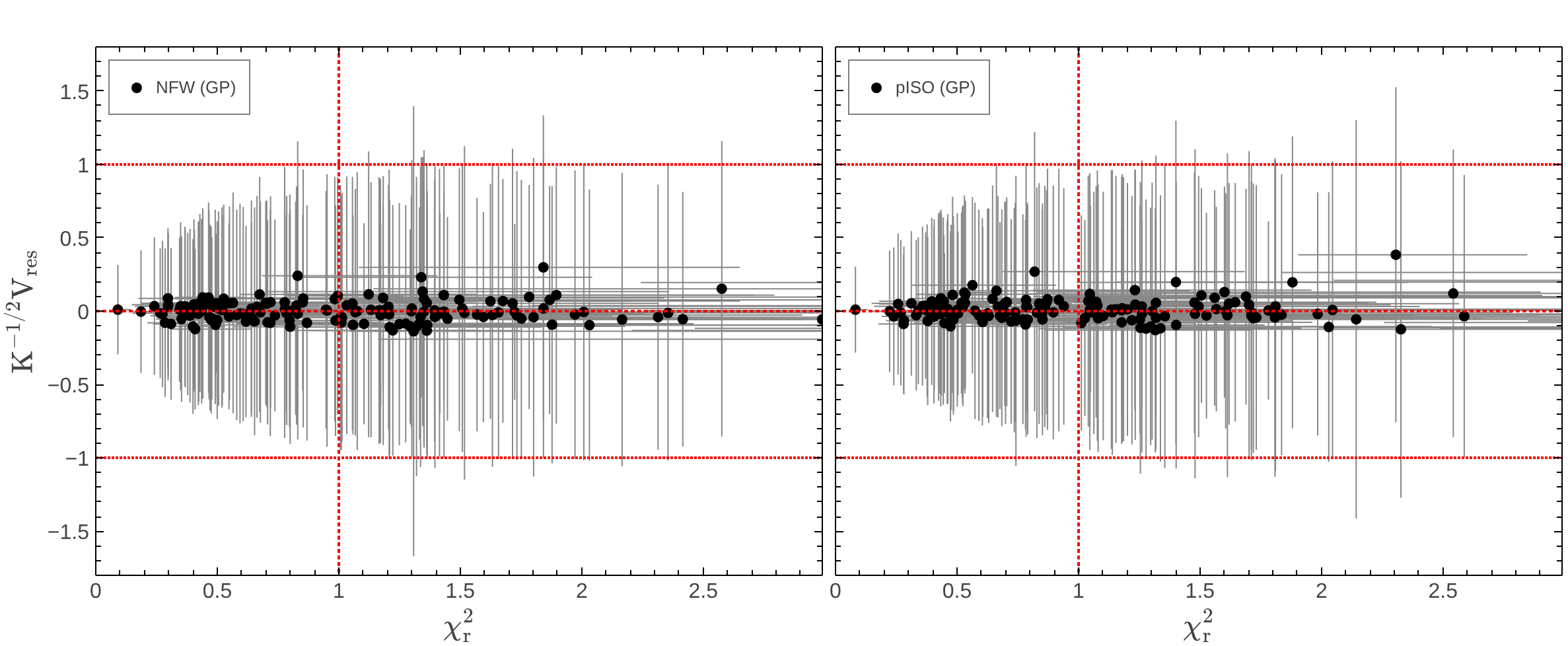}
  \caption{Comparison of goodness-of-fit statistics. For each galaxy in the full SPARC sample a point marks the location of the median of the distribution of the $\chi^2_\mathrm{r}$ statistic for the NFW (GP) model (left panel) and pISO (GP) model (right panel), and the mean of the distribution of whitened residuals $\mathrm{K}^{-1/2}\mathrm{V}_\mathrm{res}$ for the same model. Horizontal error bars mark the $16^\mathrm{th}$-$84^\mathrm{th}$ percentile interval of the distribution of the $\chi^2_\mathrm{r}$ statistic, while vertical error bars mark the standard deviation of $\mathrm{K}^{-1/2}\mathrm{V}_\mathrm{res}$. For $\chi^2_\mathrm{r}\lesssim 1$, the standard deviation of $\mathrm{K}^{-1/2}\mathrm{V}_\mathrm{res}$ is strongly correlated with $\chi^2_\mathrm{r}$, while for larger values it approximately saturates at $1$. The horizontal broken lines mark the expected mean ($0$) and standard deviation ($\pm 1$) of the distribution of $\mathrm{K}^{-1/2}\mathrm{V}_\mathrm{res}$ for a good fit, while the vertical broken line marks $\chi^2_\mathrm{r}=1$, also the expectation for a good fit. This provides reassurance that $\chi^2_\mathrm{r}$ is a useful goodness-of-fit statistic for our purposes.}
  \label{fig:whitened}
\end{figure*}

Fig.~\ref{fig:whitened} shows the mean and standard deviation of the $\mathrm{K}^{-1/2}\mathrm{V}_\mathrm{res}$ distribution for each galaxy against the median and $16^\mathrm{th}$-$84^\mathrm{th}$ percentiles of its $\chi^2_\mathrm{r}$ distribution. The mean of $\mathrm{K}^{-1/2}\mathrm{V}_\mathrm{res}$ is usually close to $0$, while its standard deviation is related to $\chi^2_\mathrm{r}$: smaller $\chi^2_\mathrm{r}$ values correspond to smaller $\mathrm{K}^{-1/2}\mathrm{V}_\mathrm{res}$ values below $\chi^2_\mathrm{r}\sim 1$. Above this threshold the standard deviation of $\mathrm{K}^{-1/2}\mathrm{V}_\mathrm{res}$ saturates, rarely much exceeding $1$. We also investigated the skew and kurtosis of the $\mathrm{K}^{-1/2}\mathrm{V}_\mathrm{res}$ distribution as possible metrics, but the correspondence between the standard deviation and $\chi^2_\mathrm{r}$ is stronger. Taking this together with considering visualisations similar to Fig.~\ref{fig:DDO154-full} we conclude that $\chi^2_\mathrm{r}$ is a useful statistic for assessing goodness of fit for our purposes and therefore use it as our fiducial statistic for this purpose throughout our analysis.

\section{DDO~154: marginalized posterior probability distributions} \label{appendix:corner}

\begin{figure*}
    \centering
    \includegraphics[width=\linewidth]{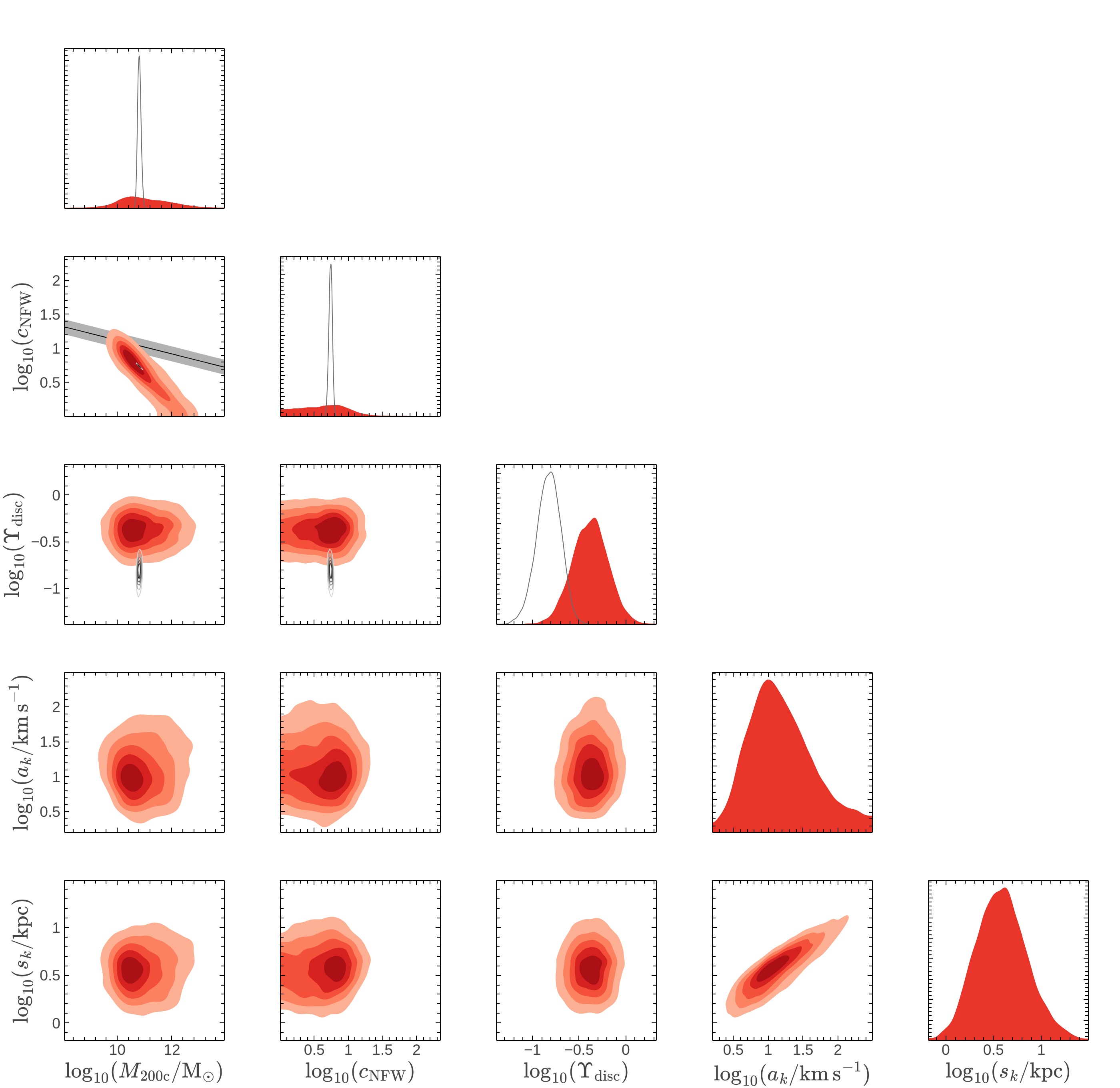}
    \caption{Corner plot showing the one- (diagonal panels) and two-dimensional (off-diagonal panels) marginalized posterior probability distributions of the parameters of our nGP (black lines and greyscale contours) and GP (red contours and filled curves) models using the NFW dark halo model. The nGP models do not include the $a_k$ and $s_k$ parameters. The halo mass-concentration relation of \citep{2014MNRAS.441.3359D} and its scatter is shown in the $M_\mathrm{200c}$-$c_\mathrm{NFW}$ panel as a black line.}
    \label{fig:NFW-corner-plot}
\end{figure*}

\begin{figure*}
    \centering
    \includegraphics[width=\linewidth]{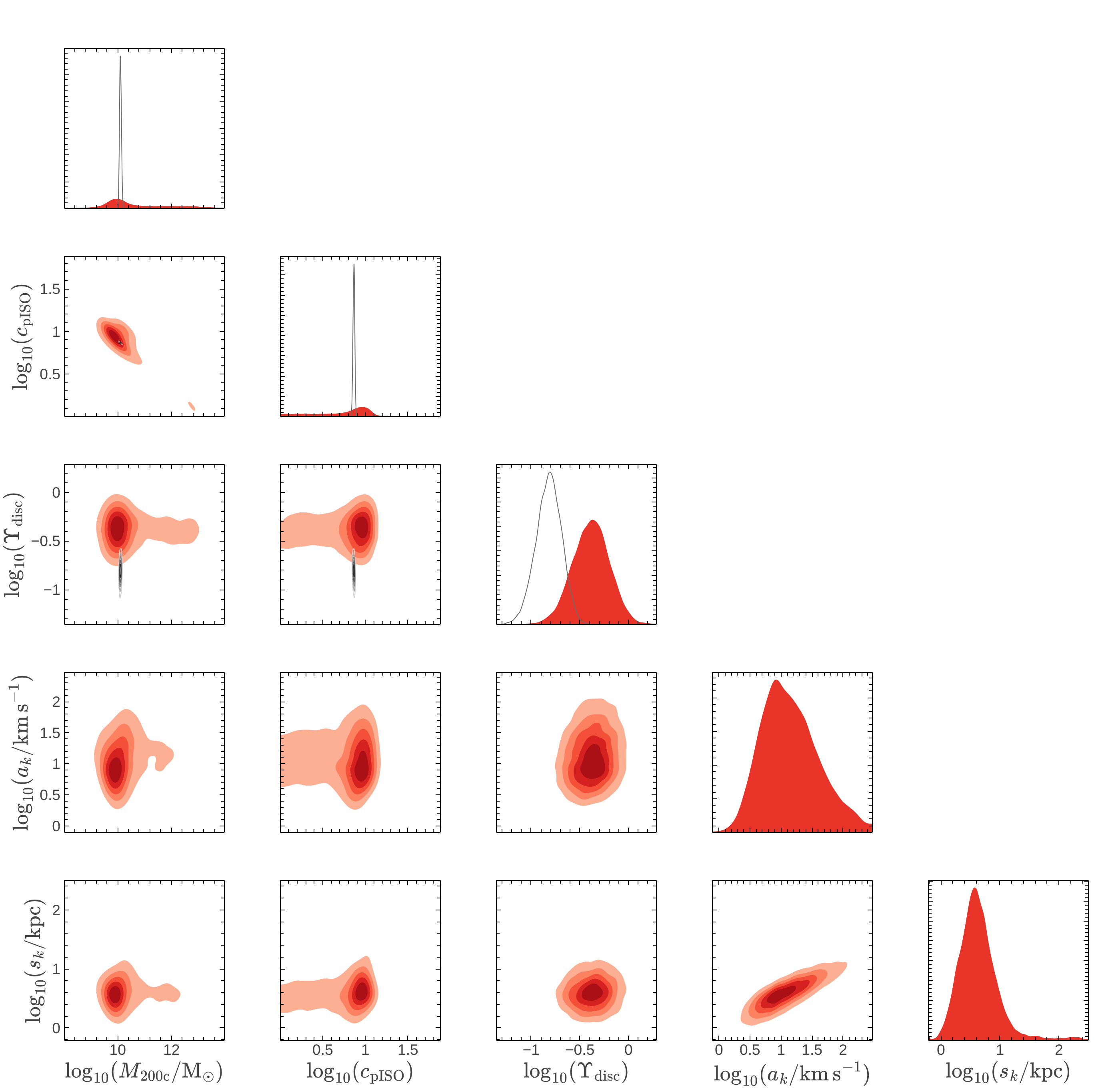}
    \caption{As Fig.~\ref{fig:NFW-corner-plot} but for the pISO halo model. There is no analogue to the halo mass-concentration relation for the pISO dark halo model.}
    \label{fig:pISO-corner-plot}
\end{figure*}

The one- and two-dimensional marginalized posterior probability distributions for our nGP and GP model fits to galaxy DDO~154 are shown in Figs.~\ref{fig:NFW-corner-plot} \& \ref{fig:pISO-corner-plot} for the NFW and pISO halo models, respectively.

\section{Robustness against variations in methodology} \label{appendix:robust}

\subsection{Influence of the choice of kernel function} \label{appendix:kernel}

\begin{figure}
    \centering
    \includegraphics[width=\columnwidth]{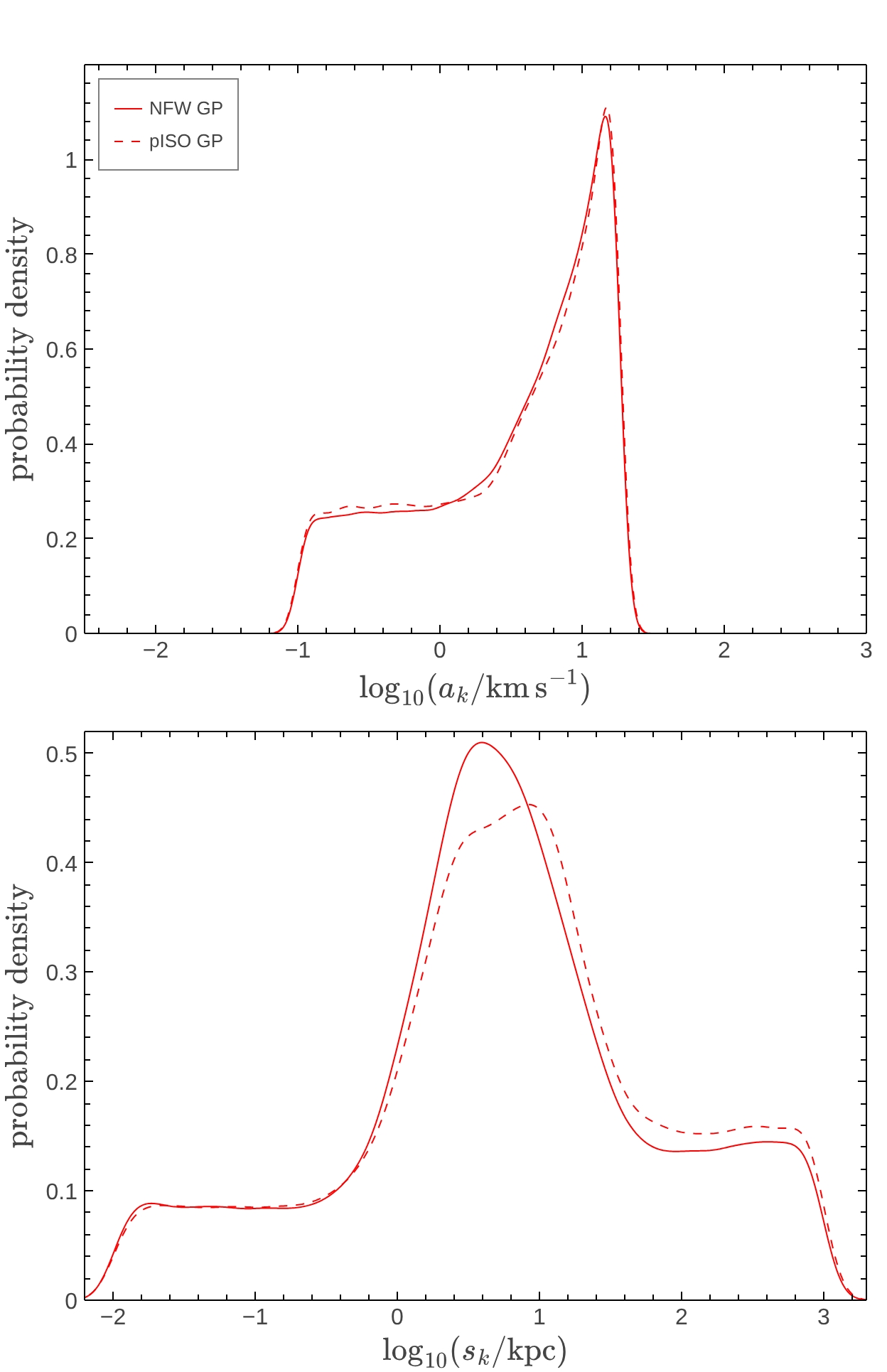}
    \caption{As Fig.~\ref{fig:ak-sk}, but using a Mat\'{e}rn-$\frac{3}{2}$ kernel to construct the covariance matrix rather than a Gaussian kernel. The parameter constraints at the population level are similar for the two kernels, except that the models with the Mat\'{e}rn-$\frac{3}{2}$ kernel lack the tails at $\log_{10}(a_k/\mathrm{km}\,\mathrm{s}^{-1}) \gtrsim 1.4$ and $\log_{10}(a_k/\mathrm{km}\,\mathrm{s}^{-1}) \lesssim -1$ that occurred when the Gaussian kernel was used.}
    \label{fig:ak-sk-matern}
\end{figure}

Our main analysis was carried out using a Gaussian kernel parametrization for the covariance matrix. We have repeated our model fits using a Mat\'{e}rn-$\frac{3}{2}$ kernel instead. Individual model fits using the two kernels sometimes differ appreciably, but in most cases the differences are small. We reach the same overall conclusions using the Mat\'{e}rn-$\frac{3}{2}$ kernel as with the Gaussian kernel, in particular the lower panels of Fig.~\ref{fig:main-results} are barely distinguishable. We include a version of Fig.~\ref{fig:ak-sk} for this alternate kernel choice in Fig.~\ref{fig:ak-sk-matern}. The constraints on the $a_k$ and $s_k$ parameters across the galaxy sample are similar to with the Gaussian kernel, but the correlation amplitude is more sharply constrained to be $0.1<a_k/\mathrm{km}\,\mathrm{s}^{-1}<25$, whereas with the Gaussian kernel there was a tail extending to $>100\,\mathrm{km}\,\mathrm{s}{-1}$ -- in this sense the outcome with the Mat\'{e}rn-$\frac{3}{2}$ kernel is perhaps even more physically plausible.

\subsection{Influence of priors} \label{appendix:priors}

We experimented with alternative choices of priors including:
\begin{itemize}
\item varying the bounds on the $a_k$ and $s_k$;
\item adopting non-uniform (but still broad) priors on $a_k$ and $s_k$;
\item adopting the posterior probability distribution of the nGP model fits as priors on the $M_\mathrm{200c}$, $c$ and $\Upsilon$ parameters for the GP model fits;
\item adopting the \citet{2014MNRAS.441.3359D} halo mass-concentration relation as a prior on $c_\mathrm{NFW}$ (this particular choice raises the question of what, if anything, to impose for $c_\mathrm{pISO}$);
\item combinations thereof.
\end{itemize}
We concluded that all of these choices can lead to situations where a model that would 
not converge to a physically plausible solution (e.g. preferring $s_k$ much larger than the disc) with uniform priors would settle to a compromise between the prior and the area of parameter space otherwise favoured. However, this makes identifying when no physically plausible solution exists more difficult. Since this likely signals a fundamental incompatibility between the halo model(s) and observed rotation curve, we found that allowing the model to fail `catastrophically' in these cases by adopting wide, uniform priors (except on the $\Upsilon$ parameters) led to the most readily interpreted posterior probability distributions. None of our experiments with alternative priors led to a clear preference for one halo model over the other.

\section{Goodness of fit for all models} \label{appendix:table}

Representative $\chi^2_\mathrm{r}$ values for nGP and GP model fits using the NFW and pISO halo models for all galaxies in our sample are tabulated in Table~\ref{tab:all-chi2}.

\onecolumn
\begin{longtable}{lccccclcccc}
  \caption{Goodness of fit for our four mass models (NFW nGP; pISO nGP; NFW GP; pISO GP) for all galaxies in our subsample from SPARC. The values are the medians of the distribution of $\chi^2_\mathrm{r}$ values across MCMC samples.}\\
  \cline{1-5}\cline{7-11}
  \rule{0pt}{3ex} % avoid exponent touching line
  &\multicolumn{4}{c}{$\chi^2_\mathrm{r}$} && &\multicolumn{4}{c}{$\chi^2_\mathrm{r}$} \\
  \cline{2-5} \cline{8-11}
  Name & NFW nGP & pISO nGP & NFW GP & pISO GP && Name & NFW nGP & pISO nGP & NFW GP & pISO GP \\
  \cline{1-5}\cline{7-11}
  \endfirsthead
  \cline{1-5}\cline{7-11}
  \rule{0pt}{3ex} % avoid exponent touching line
  &\multicolumn{4}{c}{$\chi^2_\mathrm{r}$} && &\multicolumn{4}{c}{$\chi^2_\mathrm{r}$} \\
  \cline{2-5} \cline{8-11}
  Name & NFW nGP & pISO nGP & NFW GP & pISO GP && Name & NFW nGP & pISO nGP & NFW GP & pISO GP \\
  \cline{1-5}\cline{7-11}
  \endhead
  \cline{1-5}\cline{7-11} \multicolumn{11}{r}{{\emph{Continued on next page.}}} \\
  \endfoot
  \cline{1-5}\cline{7-11}
  \endlastfoot
  D631-7 & 5.1 & 4.9 & 0.6 & 0.6    &&  NGC~5985 & 2.0 & 3.4 & 0.6 & 0.6 \\ 
  DDO~064 & 0.7 & 0.7 & 0.8 & 0.8    &&  NGC~6015 & 8.3 & 8.2 & 0.8 & 0.9 \\ 
  DDO~154 & 10.1 & 6.7 & 1.2 & 1.3    &&  NGC~6195 & 1.9 & 9.1 & 1.4 & 1.4 \\ 
  DDO~161 & 0.9 & 0.6 & 0.4 & 0.4    &&  NGC~6503 & 1.5 & 2.5 & 1.2 & 1.1 \\ 
  DDO~170 & 2.9 & 2.7 & 4.5 & 4.2    &&  NGC~6674 & 2.1 & 1.7 & 1.5 & 1.9 \\ 
  ESO079-G014 & 4.0 & 3.8 & 1.0 & 1.0    &&  NGC~6946 & 1.7 & 1.5 & 0.4 & 0.4 \\ 
  ESO116-G012 & 2.7 & 1.7 & 1.0 & 1.1    &&  NGC~7331 & 0.9 & 1.1 & 0.4 & 0.4 \\ 
  ESO563-G021 & 17.7 & 17.7 & 1.2 & 1.2    &&  NGC~7793 & 1.0 & 0.8 & 0.4 & 0.3 \\ 
  F563-1 & 1.3 & 1.1 & 1.3 & 1.2    &&  NGC~7814 & 0.6 & 0.9 & 0.7 & 0.9 \\ 
  F563-V2 & 1.5 & 1.4 & 1.9 & 1.8    &&  UGC~00128 & 3.3 & 5.7 & 1.2 & 1.1 \\ 
  F568-1 & 1.1 & 1.1 & 1.3 & 1.3    &&  UGC~00191 & 3.7 & 3.6 & 3.6 & 3.2 \\ 
  F568-3 & 3.5 & 3.4 & 0.5 & 0.5    &&  UGC~00731 & 0.5 & 0.3 & 0.6 & 0.3 \\ 
  F568-V1 & 0.4 & 0.2 & 0.4 & 0.3    &&  UGC~01230 & 1.4 & 1.0 & 1.7 & 1.2 \\ 
  F571-8 & 3.2 & 2.3 & 1.0 & 1.1    &&  UGC~02259 & 1.2 & 2.5 & 1.9 & 2.2 \\ 
  F574-1 & 1.9 & 1.3 & 0.6 & 0.6    &&  UGC~02487 & 4.4 & 4.5 & 0.8 & 0.8 \\ 
  F579-V1 & 0.3 & 0.4 & 0.4 & 0.5    &&  UGC~02885 & 1.0 & 26.2 & 1.2 & 1.3 \\ 
  F583-1 & 1.7 & 1.4 & 0.4 & 0.4    &&  UGC~02916 & 10.8 & 1.8 & 0.3 & 0.4 \\ 
  F583-4 & 0.3 & 0.3 & 0.3 & 0.4    &&  UGC~02953 & 5.6 & 6.1 & 0.2 & 0.2 \\ 
  IC~2574 & 36.3 & 33.5 & 0.6 & 0.6    &&  UGC~03205 & 2.8 & 3.5 & 0.3 & 0.4 \\ 
  IC~4202 & 18.3 & 12.6 & 0.8 & 0.8    &&  UGC~03546 & 1.0 & 0.8 & 0.8 & 0.8 \\ 
  KK98-251 & 2.1 & 2.0 & 0.7 & 0.7    &&  UGC~03580 & 2.2 & 2.3 & 0.6 & 0.6 \\ 
  NGC~0024 & 0.4 & 0.4 & 0.5 & 0.4    &&  UGC~04278 & 1.4 & 1.3 & 0.4 & 0.4 \\ 
  NGC~0055 & 2.3 & 1.9 & 0.4 & 0.4    &&  UGC~04325 & 3.0 & 2.1 & 2.1 & 3.0 \\ 
  NGC~0100 & 0.9 & 0.9 & 0.4 & 0.4    &&  UGC~04483 & 1.0 & 1.0 & 1.6 & 1.6 \\ 
  NGC~0247 & 1.8 & 1.8 & 0.4 & 0.4    &&  UGC~04499 & 1.6 & 1.2 & 2.1 & 1.7 \\ 
  NGC~0289 & 2.0 & 2.0 & 0.7 & 0.6    &&  UGC~05005 & 0.4 & 0.4 & 0.5 & 0.5 \\ 
  NGC~0300 & 0.8 & 0.7 & 0.8 & 0.7    &&  UGC~05253 & 2.4 & 0.8 & 0.4 & 0.4 \\ 
  NGC~0801 & 5.6 & 5.6 & 3.7 & 3.6    &&  UGC~05716 & 2.1 & 1.8 & 1.5 & 2.0 \\ 
  NGC~0891 & 2.5 & 1.3 & 1.2 & 1.1    &&  UGC~05721 & 1.0 & 0.6 & 0.8 & 0.6 \\ 
  NGC~1003 & 2.4 & 2.5 & 0.5 & 0.5    &&  UGC~05750 & 1.2 & 1.1 & 1.4 & 1.3 \\ 
  NGC~1090 & 2.6 & 2.1 & 1.2 & 0.9    &&  UGC~05764 & 7.9 & 4.7 & 2.0 & 2.5 \\ 
  NGC~1705 & 0.2 & 0.5 & 0.3 & 0.5    &&  UGC~05829 & 0.2 & 0.2 & 0.3 & 0.2 \\ 
  NGC~2366 & 3.3 & 2.8 & 0.4 & 0.4    &&  UGC~05918 & 0.5 & 0.7 & 0.9 & 1.1 \\ 
  NGC~2403 & 9.2 & 11.4 & 0.6 & 0.6    &&  UGC~05986 & 5.7 & 5.2 & 0.5 & 0.5 \\ 
  NGC~2683 & 1.4 & 1.2 & 1.9 & 1.7    &&  UGC~06399 & 0.9 & 0.9 & 1.3 & 1.3 \\ 
  NGC~2841 & 1.5 & 1.7 & 0.4 & 0.5    &&  UGC~06446 & 0.3 & 0.4 & 0.4 & 0.4 \\ 
  NGC~2903 & 6.0 & 7.0 & 1.3 & 1.0    &&  UGC~06614 & 0.6 & 0.5 & 0.7 & 0.6 \\ 
  NGC~2915 & 0.9 & 0.7 & 0.5 & 0.5    &&  UGC~06667 & 1.8 & 1.3 & 2.3 & 1.7 \\ 
  NGC~2955 & 3.5 & 6.8 & 1.4 & 1.4    &&  UGC~06786 & 0.7 & 1.5 & 0.7 & 0.6 \\ 
  NGC~2976 & 0.9 & 0.9 & 0.5 & 0.5    &&  UGC~06787 & 18.3 & 18.2 & 0.6 & 0.6 \\ 
  NGC~2998 & 2.2 & 3.0 & 1.1 & 1.1    &&  UGC~06917 & 1.1 & 0.8 & 1.4 & 1.0 \\ 
  NGC~3109 & 9.1 & 8.7 & 0.6 & 0.5    &&  UGC~06930 & 0.6 & 0.6 & 0.8 & 0.8 \\ 
  NGC~3198 & 1.4 & 1.4 & 0.4 & 0.4    &&  UGC~06973 & 1.4 & 1.6 & 2.2 & 2.4 \\ 
  NGC~3521 & 0.3 & 0.3 & 0.3 & 0.3    &&  UGC~06983 & 0.8 & 0.6 & 0.9 & 0.7 \\ 
  NGC~3726 & 2.4 & 2.3 & 1.8 & 1.9    &&  UGC~07089 & 0.4 & 0.3 & 0.5 & 0.4 \\ 
  NGC~3741 & 0.5 & 0.5 & 0.5 & 0.5    &&  UGC~07125 & 1.1 & 0.7 & 1.2 & 0.8 \\ 
  NGC~3769 & 0.8 & 1.0 & 1.0 & 1.2    &&  UGC~07151 & 2.5 & 2.3 & 1.4 & 1.5 \\ 
  NGC~3893 & 0.9 & 0.6 & 1.2 & 0.8    &&  UGC~07323 & 0.9 & 0.8 & 1.1 & 1.0 \\ 
  NGC~3917 & 3.0 & 2.9 & 0.8 & 0.8    &&  UGC~07399 & 1.4 & 1.1 & 1.7 & 1.5 \\ 
  NGC~3953 & 1.0 & 1.0 & 1.6 & 1.4    &&  UGC~07524 & 0.9 & 0.6 & 0.3 & 0.4 \\ 
  NGC~3972 & 1.5 & 1.5 & 2.0 & 2.0    &&  UGC~07577 & 0.4 & 0.3 & 0.5 & 0.5 \\ 
  NGC~3992 & 0.9 & 1.3 & 1.3 & 1.5    &&  UGC~07603 & 1.9 & 1.4 & 0.9 & 0.9 \\ 
  NGC~4010 & 2.5 & 2.5 & 1.0 & 1.1    &&  UGC~07608 & 0.7 & 0.6 & 1.1 & 1.0 \\ 
  NGC~4013 & 0.8 & 0.8 & 0.8 & 0.8    &&  UGC~08286 & 2.0 & 2.2 & 0.8 & 0.7 \\ 
  NGC~4088 & 0.6 & 0.6 & 0.8 & 0.8    &&  UGC~08490 & 0.2 & 0.2 & 0.2 & 0.3 \\ 
  NGC~4100 & 1.0 & 0.6 & 0.6 & 0.6    &&  UGC~08550 & 0.9 & 0.8 & 1.2 & 1.1 \\ 
  NGC~4157 & 0.5 & 0.5 & 0.6 & 0.6    &&  UGC~08699 & 0.7 & 0.8 & 0.7 & 0.7 \\ 
  NGC~4183 & 0.3 & 0.4 & 0.3 & 0.4    &&  UGC~09037 & 1.5 & 1.2 & 0.7 & 0.8 \\ 
  NGC~4214 & 0.9 & 1.0 & 0.9 & 0.8    &&  UGC~09133 & 7.0 & 7.4 & 0.7 & 0.6 \\ 
  NGC~4217 & 2.1 & 2.0 & 0.5 & 0.5    &&  UGC~11455 & 5.1 & 5.0 & 0.8 & 0.9 \\ 
  NGC~4559 & 0.3 & 0.2 & 0.3 & 0.2    &&  UGC~11820 & 0.9 & 0.8 & 1.3 & 1.2 \\ 
  NGC~5005 & 0.2 & 0.3 & 0.3 & 0.3    &&  UGC~11914 & 0.9 & 7.5 & 0.2 & 0.2 \\ 
  NGC~5033 & 2.9 & 2.0 & 0.9 & 0.8    &&  UGC~12506 & 0.2 & 0.5 & 0.2 & 0.3 \\ 
  NGC~5055 & 2.7 & 2.6 & 1.1 & 1.0    &&  UGC~12632 & 0.5 & 0.3 & 0.5 & 0.3 \\ 
  NGC~5371 & 4.4 & 10.7 & 1.2 & 1.2    &&  UGC~12732 & 0.3 & 0.5 & 0.4 & 0.5 \\ 
  NGC~5585 & 6.0 & 5.3 & 1.2 & 1.2    &&  UGCA~442 & 2.8 & 2.2 & 2.8 & 2.8 \\ 
  NGC~5907 & 4.3 & 5.3 & 1.4 & 1.4    &&  UGCA~444 & 0.1 & 0.1 & 0.1 & 0.1 \\    
  \label{tab:all-chi2}
\end{longtable}

\bsp
\label{lastpage}
\end{document}